\documentclass[11pt, a4paper, onecolumn, copyright, gr]{google}

\usepackage[authoryear, compress, round]{natbib}
\uselogo

\usepackage{microtype}
\usepackage{graphicx}
\usepackage{subcaption}
\usepackage{booktabs}

\usepackage{amsmath}
\usepackage{amssymb}
\usepackage{mathtools}
\usepackage{amsthm}

\usepackage{titletoc}
\usepackage{hyperref}  
\usepackage[page,header]{appendix}

\newcommand{\runningTitle}{Discovering Mechanistic Models of Neural Activity}
\newcommand{\longTitle}{\runningTitle: System Identification in an \textit{in Silico} Zebrafish}

\title{\longTitle}

\usepackage[capitalize,noabbrev,nameinlink]{cleveref}
\creflabelformat{equation}{#2#1#3}

\usepackage{multibib}
\newcites{app}{References}

\theoremstyle{plain}

\theoremstyle{definition}

\theoremstyle{remark}

\usepackage{listings}
\usepackage{xcolor}
\usepackage{inconsolata}
\usepackage[most]{tcolorbox}

\tcbuselibrary{listings, breakable}

\definecolor{atompurple}{RGB}{166, 38, 164}
\definecolor{atomblue}{RGB}{64, 120, 242}
\definecolor{atomcyan}{RGB}{1, 132, 188}
\definecolor{atomgreen}{RGB}{80, 161, 79}
\definecolor{atomred}{RGB}{228, 86, 73}
\definecolor{atomgray}{RGB}{160, 161, 167}
\definecolor{bglight}{RGB}{250, 250, 250}

\lstdefinestyle{pythonlst}{
    language=Python,
    backgroundcolor=\color{bglight},
    commentstyle=\color{atomgray}\itshape,
    keywordstyle=\color{atompurple}\bfseries,
    stringstyle=\color{atomgreen},
    numberstyle=\ttfamily\tiny\color{atomgray},
    basicstyle=\ttfamily\scriptsize\linespread{1.1},
    morekeywords={self, True, False, None, as, with, yield},
    emph={jnp, nn, flax, optax, jax, numpy, plt},
    emphstyle=\color{atomcyan}\bfseries,
    classoffset=1, 
    morekeywords={def, class},
    keywordstyle=\color{atomblue}\bfseries,
    classoffset=0,
    showstringspaces=false,
    captionpos=b,
    frame=none, 
    belowskip=0em,
    aboveskip=0em
}

\newtcblisting{pythonbox}{
    enhanced,
    boxrule=0.5pt,
    colback=bglight,
    colframe=gray!30, 
    breakable,
    top=4mm, bottom=4mm, left=8mm, right=4mm,
    listing only,
    listing options={
        style=pythonlst,
        basicstyle=\ttfamily\scriptsize,
        numbers=left,
        breaklines=true,
        columns=fullflexible
    }
}

\lstdefinestyle{markdownlst}{
    language={},
    backgroundcolor=\color{bglight},
    basicstyle=\ttfamily\scriptsize\linespread{1.1},
    commentstyle=\color{atompurple}\bfseries, 
    morecomment=[l]{\#},
    stringstyle=\color{atomgreen},
    morestring=[b]`,
    showstringspaces=false,
    breaklines=true,
    breakatwhitespace=true,
    columns=fullflexible,
    frame=none,
    aboveskip=0em,
    belowskip=0em
}

\newtcblisting{promptbox}{
    enhanced,
    boxrule=0.5pt,
    colback=bglight,
    colframe=gray!30,
    breakable,
    top=4mm, bottom=4mm, left=4mm, right=4mm,
    listing only,
    listing options={
        style=markdownlst,
        numbers=none,
    }
}

\makeatletter
\newcommand{\standardaddcontentsline}[3]{%
  \addtocontents{#1}{\protect\contentsline{#2}{#3}{\thepage}{\@currentHref}}%
}
\makeatother

\newcommand{\capskip}{\vskip -1em}

\NewDocumentCommand{\model}{m o}{%
  \texttt{#1}%
  \IfValueT{#2}{\textsubscript{\texttt{#2}}}%
}

\reportnumber{} 

\author[1]{Jan-Matthis Lueckmann}
\author[1]{Viren Jain}
\author[1]{Michał Januszewski}

\affil[1]{\thepa{}{}}

\begin{abstract}
Constructing mechanistic models of neural circuits is a fundamental goal of neuroscience, yet verifying such models is limited by the lack of ground truth.
To rigorously test model discovery, we establish an \textit{in silico} testbed using neuromechanical simulations of a larval zebrafish as a transparent ground truth. 
We find that LLM-based tree search autonomously discovers predictive models that significantly outperform established forecasting baselines. 
Conditioning on sensory drive is necessary but not sufficient for faithful system identification, as models exploit statistical shortcuts.
Structural priors prove essential for enabling robust out-of-distribution generalization and recovery of interpretable mechanistic models.
Our insights provide guidance for modeling real-world neural recordings and offer a broader template for AI-driven scientific discovery.
\end{abstract}

\correspondingauthor{janmatthis@google.com, mjanusz@google.com}

\begin{document}

\maketitle

\section{Introduction}

A fundamental goal in neuroscience is to understand how neural circuits process information and generate behavior. The ability to predict neural activity is a key test towards such understanding. Data-driven benchmarks can accelerate progress by providing standardized datasets and evaluation metrics \citep{schrimpf2020integrative,pei2022neural,turishcheva2024dynamic,lueckmann2025zapbench}.

However, benchmarks relying on real experimental data face inherent limitations and verification challenges. For instance, limitations include the high cost and technical difficulty of acquiring large-scale, high-resolution neural recordings, the presence of noise and unobserved variables, and restrictions in systematically manipulating neural circuits or environments. Real-world experiments that involve imaging of neural activity are severely restricted in duration and throughput (e.g., due to phototoxicity, molecular reporters, and imaging limitations), and subject to artifacts and preprocessing challenges (e.g., motion, striping, imperfect alignment and segmentation). More fundamentally, much about the underlying biological systems used in such benchmarks is as yet unknown (e.g., the detailed connectivity of underlying circuits, synaptic release probabilities, neuromodulatory interactions). Therefore, since the ground-truth data-generating process is unknown, models are ranked relative to each other rather than compared against an absolute reference. It is thus unclear how far models are from the true mechanisms of computation, or whether they have identified statistical shortcuts. This creates a verification gap: without a known ground truth, it is hard to rigorously evaluate models and model discovery strategies. 

Neuromechanical simulations provide a powerful avenue to complement real-data benchmarks: Embodied simulations can generate vast amounts of perfectly annotated data where the underlying ``ground truth'' of the neural system is known and controllable. They allow for systematic manipulation of physiological and circuit features that are difficult or impossible to achieve in biological experiments---offering full observability and control over factors such as dataset dimensionality, length, and noise levels, along with the ability to perturb any part of their neural circuitry. \textit{simZFish} \citep{liu2024artificial}, an open-source neuromechanical simulation of the larval zebrafish, is one such platform that realistically models the fish's body, its physical interactions with the environment, visual inputs, and experimentally constrained neural architectures for visuomotor behaviors. 

\begin{figure*}[t]
  \begin{center}
    \centerline{\includegraphics[width=\linewidth]{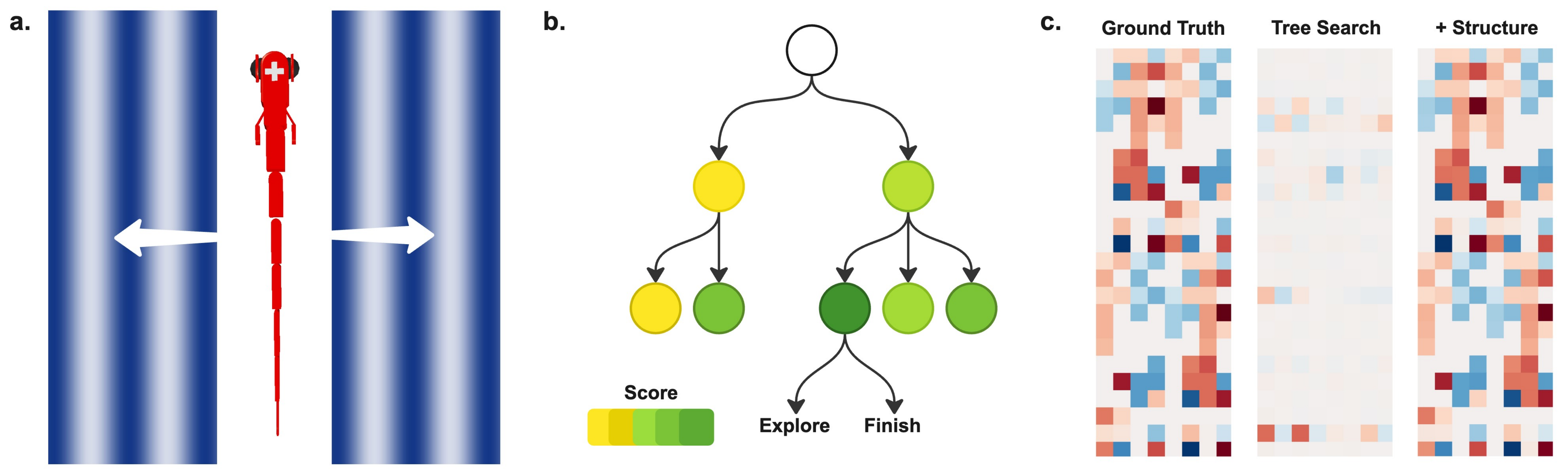}}
    \caption{
      \textbf{ Verifiable discovery of neural mechanisms in an \textit{in silico} testbed.}
      \textbf{a.} The simulation environment consists of a neuromechanical model subject to fluid dynamics, responding to visual stimuli driven by a neural circuit in a closed-loop setting.
      \textbf{b.} We use LLM-guided tree search to autonomously explore the space of dynamical models, evolving Python code to minimize predictive error on neural activity.
      \textbf{c.} Despite high predictive in-distribution performance, unconstrained black-box tree search models fail to identify the system's mechanisms, as revealed by effective connectivity matrices (excerpt shown; blue and red indicate inhibitory and excitatory interactions, respectively; color intensity represents magnitude). In contrast, a structure-constrained grey-box tree search model successfully identifies the correct signs and magnitudes closely matching the ground truth, from a structural prior that provides information about the existence and absence of connections.
    }
    \label{fig:overview}
    \phantomsubcaption\label{fig:overview-a}
    \phantomsubcaption\label{fig:overview-b}
    \phantomsubcaption\label{fig:overview-c}
  \end{center}
  \capskip
\end{figure*}

In this work, we use simZFish to generate high-fidelity \textit{in silico} datasets that serve as a testbed for system identification \citep{ljung2010perspectives}. By simulating the virtual fish's response to a range of visual stimuli, we create a ground-truth dataset where every internal state, sensory input, and underlying circuit connectivity is perfectly observable. This allows us to study a critical question: What are the data and modeling prerequisites to recover the mechanisms of a complex neural circuit?

While standard forecasting architectures can provide initial baselines, they are rarely optimized for the specific temporal and structural constraints of neural dynamics. A primary bottleneck in modeling neural activity is often the modeler's own inductive bias and the significant time required to manually explore the architectural landscape. We address this by framing model design as an optimization problem. Specifically, we utilize an automated tree-search approach \citep{aygun2025ai} to traverse a vast space of code-based model definitions, leveraging Large Language Models (LLMs) to effectively ``evolve'' solutions. We compare two distinct discovery regimes: an unconstrained search for high-accuracy black-box predictors, and a structure-constrained search that uses structural priors (derived from the ``wiring diagram''), and we evaluate the discovered models with respect to their ability to generalize to unseen stimuli, their interpretability, and mechanistic recovery~(\cref{fig:overview}).

In summary, our contributions are as follows:
\begin{enumerate}
\item \textbf{An \textit{in silico} testbed for verifiable neural system identification.} Unlike real-data benchmarks where ground truth is unknown, our testbed provides a transparent reference, allowing us to evaluate whether models recover the true mechanisms of a neural circuit.
\item \textbf{Sensory drive is a prerequisite for identifiability.} We demonstrate that in sensory-driven neural circuits, the true transition function is practically non-identifiable using standard autoregressive history alone.
\item \textbf{In-distribution accuracy is a poor proxy for system identification.} We show that LLM-guided tree search discovers models that achieve state-of-the-art test set accuracy but fail on out-of-distribution (OOD) stimuli. These models prioritize statistical shortcuts over faithful recovery of mechanisms, challenging the practice of ranking neural models by test set error.
\item \textbf{Structural priors enable mechanistic recovery.} We demonstrate that providing structural priors is sufficient to regularize the discovery process. This eliminates statistical shortcuts, enables robust OOD generalization, interpretability, and faithful recovery of static and dynamic mechanisms---as verified by effective connectivity and impulse response analyses, respectively.
\item \textbf{Recommendations for real-world system identification:} Based on our \textit{in silico} results, we propose concrete guidelines for modeling real neural recordings. We recommend shifting from pure autoregression to connectome-constrained conditional forecasting tasks with OOD evaluations, to bridge the gap between predictive AI and mechanistic neuroscience.
\end{enumerate}

We position our work relative to recent advances in automated model discovery, neural system identification, and connectome-constrained modeling in \cref{app:related_work}.

\section{Methods}

\subsection{Neuromechanical Simulations}
\label{sec:neuromechanical_simulations}

\begin{figure*}[t]
  \begin{center}
    \centerline{\includegraphics[width=\linewidth]{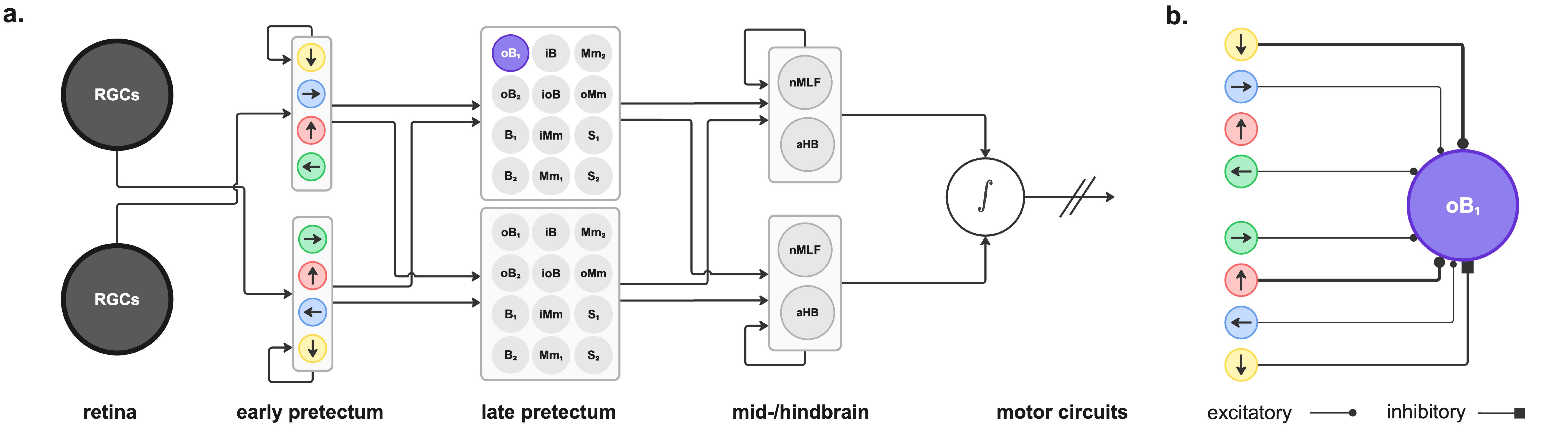}}
    \caption{
      \textbf{Neural circuit model.}
      \textbf{a.}~The neural circuit model defines the information flow, processing retinal input through the early and late pretectum (ePT, lPT) to drive downstream command nuclei (nMLF, aHB) and motor circuits.
      \textbf{b.}~Connectivity diagram of an example late pretectal neuron ($\text{oB}_\text{1}$), illustrating the specific excitatory and inhibitory wiring with ePT neurons resulting in its direction-selectivity.
    }
    \label{fig:neural_circuit}
    \phantomsubcaption\label{fig:neural_circuit-a}
    \phantomsubcaption\label{fig:neural_circuit-b}
  \end{center}
  \capskip
\end{figure*}

We generate synthetic neural activity datasets with simZFish, a detailed neuromechanical simulator of larval zebrafish \cite{liu2024artificial}. It integrates realistic body mechanics, environmental interactions, and neural circuits to reproduce sensorimotor behaviors like the optomotor response (OMR). The OMR is an innate reflex causing fish to orient in the direction of whole-field visual motion to stabilize position, and it is probed in openly available whole-brain activity recordings \citep[e.g.,][]{chen2018brain,lueckmann2025zapbench}. We provide a summary of the simulations, and refer to \citet{liu2024artificial} and \cref{app:simulations} for full details. 

\subsubsection{Simulation Environment}

Simulations are implemented in \textit{Webots} \citep{michel2004webots}, an open-source robotics simulator. An illustration of the simulation environment is in \cref{fig:overview-a}.

\textbf{Body Model.} The body approximates a 4 mm larval zebrafish with density slightly less than water. It comprises seven cuboid segments---a head with cameras and pectoral fins, followed by six body segments, and a tail fin---connected by servomotor-actuated hinge joints.

\textbf{Sensory Environment.} Visual stimuli (drifting sinusoidal gratings) are presented in a simulated petri dish, locked to the body orientation, mimicking closed-loop setups.

\textbf{Fluid Dynamics.} To simulate realistic locomotion, the body is subject to buoyancy, viscous resistance, and inertial drag, with parameters tuned to match kinetic recordings.

Our simulations employ a 0.05 ms integration time step using Webots' default first-order semi-implicit integrator for the body kinematics.

\subsubsection{Neural Circuit Model}

The neural circuit model is based on anatomical, functional, and computational findings from real zebrafish visuomotor pathways. We ported the C-based implementation by \citet{liu2024artificial} to JAX \citep{jax2018github}, for differentiability and ease of experimentation. An overview of the network model is in \cref{fig:neural_circuit}.

The circuit is formulated as a dynamical system where the activity of a neural population is determined by a non-linear transition operator. We define a generalized sigmoidal response function, $\Phi$, parameterized by synaptic weights $\mathbf{W}$, bias thresholds $\mathbf{b}$, and response gains $\boldsymbol{\omega}$:
\begin{equation}
\Phi(\mathbf{x}; \mathbf{W}, \mathbf{b}, \boldsymbol{\omega}) = \sigma \left( \boldsymbol{\omega} \odot (\mathbf{W}\mathbf{x} - \mathbf{b}) \right),
\label{eq:general_activation}
\end{equation}
where $\mathbf{x}$ is the input vector, $\sigma(\cdot)$ is the element-wise sigmoid function, and $\odot$ denotes the Hadamard product.

\textbf{Retina.} Visual input is captured by two laterally positioned simulated cameras, acting as photoreceptors driving a layer of simulated Bipolar Cells (BCs) detecting transient luminance changes. BCs project to four types of Direction-Selective Retinal Ganglion Cells (DSGCs), tuned to the four cardinal directions, producing an output vector $\mathbf{u}^{\text{ret}}_t$ representing the instantaneous motion across the visual field.

\textbf{Early Pretectum.} The retinal output projects contralaterally to the Early Pretectum (ePT). Unlike the retina, pretectal responses exhibit significant temporal persistence. The ePT state $\mathbf{h}^{\text{ept}}_t$ is defined as a first-order low-pass filter driven by sparse retinotopic projections:
\begin{equation}
\mathbf{h}^{\text{ept}}_t = (1-\alpha_t)\mathbf{h}^{\text{ept}}_{t-1} + \alpha_t \Phi(\mathbf{u}^{\text{ret}}_t; \mathbf{W}_{\text{ret}}, \mathbf{b}_{\text{ept}}, \boldsymbol{\omega}_{\text{ept}}),
\label{eq:ept_layer}
\end{equation}
where $\mathbf{W}_{\text{ret}}$ is a fixed, sparse binary mask enforcing contralateral connectivity to the lower-temporal visual field. Crucially, the update rate $\alpha_t$ is state-dependent: its value depends on the current motor state (swim bout vs. rest) to dynamically filter out self-generated visual motion artifacts during rapid tail beats.

\textbf{Late Pretectum.} The ePT neurons drive the Late Pretectum (lPT), a layer responsible for synthesizing complex binocular motion primitives. This layer consists of 12 distinct functional types per hemisphere, whose names reflect three axes of variation: (1) direction (inward/outward/global); (2) ocularity (monocular/binocular); and (3) velocity preference  (type 1/type 2; forward- vs. backward-preferring). For example, an $\text{oB}_\text{1}$ neuron responds to outward binocular motion and prefers forward velocity.

Mathematically, we model the lPT as an instantaneous layer combining linear connectivity with a non-linear binocular gating mechanism. We express this compactly by treating the gating signal as a state-dependent threshold modulation:
\begin{equation}
\mathbf{h}^{\text{lpt}}_t = \Phi \left( \mathbf{h}^{\text{ept}}_t ; \mathbf{W}_{\text{lpt}}, \mathbf{b}_{\text{lpt}} - \mathbf{M}_{\text{gate}} \odot \mathbf{g}(\mathbf{h}^{\text{ept}}_t), \boldsymbol{\omega}_{\text{lpt}} \right).
\label{eq:lpt_layer}
\end{equation}
Here, $\mathbf{W}_{\text{lpt}}$ encodes the subtype-specific excitatory/inhibitory wiring (e.g., \cref{fig:neural_circuit-b}). The term $\mathbf{g}(\cdot)$ represents a binocular co-activation signal (AND-gate) implemented via a minimum function that selects specific pairs of ipsilateral and contralateral inputs (see \cref{app:simulations} for details). This modulation effectively acts as a dynamic bias, controlled by the sparse mask $\mathbf{M}_{\text{gate}}$ (active only for two lPT neurons per hemisphere).

\textbf{Command Neurons in Mid- and Hindbrain (CMD).} The lPT layer projects to two downstream command centers: the Nucleus of the Medial Longitudinal Fasciculus (nMLF) controls swim vigor, and the Anterior Hindbrain (aHB), controlling turning bias. They are modeled as first-order low-pass filters with distinct timescales $\beta$ and connectivity:
\begin{equation}
\mathbf{h}^{\text{cmd}}_t = (1-\beta)\mathbf{h}^{\text{cmd}}_{t-1} + \beta \Phi(\mathbf{h}^{\text{lpt}}_t; \mathbf{W}_{\text{cmd}}, \mathbf{b}_{\text{cmd}}, \boldsymbol{\omega}_{\text{cmd}}).
\label{eq:command_layer}
\end{equation}
While the nMLF integrates inputs to drive forward velocity, aHB integrates bilateral differences for the turning direction.

\textbf{Motor Circuits.} Locomotion in zebrafish is discrete (burst-and-glide). The ``Bout Gate'' is implemented as a stochastic Integrate-and-Fire mechanism. An evidence accumulator $I_t$ integrates the total drive from the nMLF:
\begin{equation}
I_t = I_{t-1} + \gamma \sum_{i=1}^2 h^{\text{cmd}}_{t,i} + \epsilon_t, \quad \epsilon_t \sim \mathcal{N}(0, \sigma^2).
\label{eq:bout_accumulation}
\end{equation}
A swim bout is triggered if $I_t$ exceeds a threshold $\theta$, resetting the integrator. Upon initiation, the discrete action $a_t \in \{\text{Forward}, \text{Left}, \text{Right}\}$ is sampled from a categorical distribution determined by the command neuron states, with action probabilities $\mathbf{p}_t$:
\begin{equation}
\mathbf{p}_t = \frac{\mathbf{s}_t}{\|\mathbf{s}_t\|_1} \quad \text{with} \quad \mathbf{s}_t = \mathbf{W}_{\text{act}} \mathbf{h}^{\text{cmd}}_t + \mathbf{b}_{\text{act}}.
\label{eq:action_selection}
\end{equation}

When the ``Behavior Determinator'' initiates a turn, it triggers greater activity in the ventral spinal neurons (vSPN) on one side, which in turn leads to increased ipsilateral motor neuron activation and the necessary body curvature for the turn. Finally, twelve spinal central pattern generators (CPGs), modeled as coupled oscillators, generate rhythmic tail undulations, and twelve motor neurons determine body curvature by integrating outputs from vSPNs and CPGs.

\subsubsection{Simulated Behavior}

The simulated fish executes discrete bouts consisting of rapid tail undulations followed by a passive glide phase, similar to real zebrafish. When exposed to drifting sinusoidal gratings, it recapitulates the behavioral distribution of swims and turns observed in real zebrafish.

\subsubsection{Dataset Generation}
\label{sec:dataset}

We ran simZFish simulations under four binocular motion conditions designed to probe the visuomotor circuit: bilateral outward, bilateral inward, and two asymmetric configurations where eyes receive opposing inward/outward motion. At the same time, we recorded the activity, resulting in 400,000 samples at 1~ms resolution in total.

\subsection{System Identification}
\label{sec:benchmark}

To provide a foundation for verifiable model discovery, we establish a task using the simZFish dataset.

\subsubsection{Task Definition}
\label{sec:task_def}

We frame our task as a system identification problem \citep{ljung2010perspectives}. We assume the neural activity evolves according to a ground-truth transition function $f^*$ driven by past activity and sensory information:
\begin{equation}
\mathbf{a}_{t} = f^*(\mathbf{a}_{t-1}, \mathbf{x}_{t}),
\label{eq:sysid_def}
\end{equation}
where $\mathbf{a}_t \in \mathbb{R}^N$ is the activity of the $N=36$ neural units (comprised of ePT, lPT, and nMLF/aHB populations) and $\mathbf{x}_t \in \mathbb{R}^D$ represents the exogenous covariates. The covariates include variables that drive the system but are not targets of prediction: specifically the instantaneous sensory drive $\Phi(\mathbf{u}^{\text{ret}}_t)$ and the behavioral bout state. Because the internal bout trigger is stochastic (\cref{eq:bout_accumulation}), the future trajectory contains inherent (aleatoric) uncertainty. Conditioning on $\mathbf{x}$ resolves this, transforming the task from a probabilistic into a deterministic problem. 

We seek to discover a parameterized model $f_\theta$ that approximates the mechanisms of $f^*$. Since the ground-truth mechanisms are unknown in real-world settings, we do not optimize for mechanistic similarity directly. Instead, we employ multi-step forecasting as a proxy task. The identification task is to find the optimal pair $(f, \theta)$ that minimizes the divergence between the recursively simulated trajectory $\hat{\mathbf{a}}$ and the ground truth $\mathbf{a}$ over a horizon $H=256$:
\begin{equation}
\min_\theta \sum_{t=1}^H \| \hat{\mathbf{a}}_{t} - \mathbf{a}_{t} \| \quad \text{s.t.} \quad \hat{\mathbf{a}}_{t} = f_\theta(\hat{\mathbf{a}}_{t-1}, \mathbf{x}_{t}).
\label{eq:sysid_task}
\end{equation}
To test for both interpolation and robust extrapolation, we use three of the conditions described in \cref{sec:dataset} for the in-distribution regime (split chronologically into 70\% training, 10\% validation, and 20\% test), while the fourth is reserved as an out-of-distribution (OOD) holdout to evaluate whether models generalize.

\begin{figure*}[t]
  \begin{center}
    \centerline{\includegraphics[width=\linewidth]{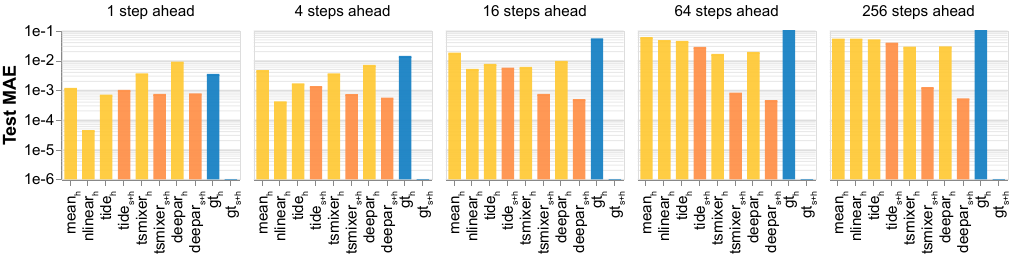}}
    \caption{
      \textbf{Sensory information is a prerequisite for system identification.} Performance (Test MAE, log scale) of baseline models and the \texttt{gt} circuit across prediction horizons. While models marked with \texttt{s+h} are conditioned on exogenous sensory drive, models with \texttt{h} only use past history. \model{gt}[h] yields higher error than the naive \model{mean}[h] baseline, suggesting that the true solution is non-identifiable without sensory drive based on standard error metrics.}
    \label{fig:sensory}
  \end{center}
  \capskip
\end{figure*}

\subsubsection{Evaluation}

We evaluate candidate models on two axes: their ability to predict neural activity and their success in recovering ground-truth mechanisms.

Following ZAPBench \citep{lueckmann2025zapbench}, we quantify trajectory prediction error using the Mean Absolute Error (MAE) averaged across all units and time steps. For a set of test windows $\mathcal{W}$, the error is defined as:
\begin{equation}
\text{MAE}_{\mathcal{W}} = \frac{1}{|\mathcal{W}|} \sum_{w \in \mathcal{W}} \frac{1}{N} \sum_{n=1}^{N} \frac{1}{H} \sum_{h=1}^{H} |\hat{a}^w_{h,n} - a^w_{h,n}|.
\label{eq:mae_metric_bench}
\end{equation}
We report this metric on two data splits. The Test MAE measures performance on held-out trajectories from the same stimulus conditions used during training. The Holdout MAE measures performance on the OOD stimulus condition. We posit that a low Holdout MAE is a strong proxy for having learned the correct mechanisms.

Since we operate in an \textit{in silico} setting with known, differentiable ground truth $f^*$, we can verify the mechanistic fidelity of a discovered model $f_\theta$: We compute the effective connectivity matrix using Jacobian sensitivity analysis and impulse responses (IR) relative to $f^*$. This allows us to evaluate recovery of static and dynamic mechanisms, respectively. \cref{app:additional_analyses} provides details on these analyses including computation of the error metrics $\mathcal{L}_{\text{Jac}}$ and $\mathcal{L}_{\text{IR}}$.

\subsection{Baseline Models}
\label{sec:baselines}

To contextualize the performance of our discovered models, we compare against two distinct categories of baselines. First, to establish the predictive ceiling of the dataset, we employ state-of-the-art time-series forecasting architectures including \model{nlinear} \citep{zeng2022are}, \model{tide} \citep{das2024longterm}, and \model{tsmixer} \citep{chen2023tsmixer}, as well as a naive \model{mean} baseline. These models map a history window directly to the future horizon without explicitly modeling step-by-step transition dynamics. While they serve as rigorous controls for predictive accuracy, they do not yield an explicit transition function. Second, we evaluate \model{deepar} \citep{salinas2020deepar}, a probabilistic recurrent network. Like our discovered models, \model{deepar} recursively predicts the next step via a learned transition function. Details are in \cref{app:baselines}.

\subsection{Tree Search}
\label{sec:methods_tree_search}

To navigate the combinatorial space of potential neural architectures, we employ an automated code generation approach driven by LLMs and Tree Search: We use the method proposed by \citet{aygun2025ai}, which combines a pre-trained LLM with a heuristic search algorithm \citep[Predictor Upper Confidence Bound;][]{silver2016mastering} to iteratively refine solutions. The search scores solution using validation set MAE on in-distribution conditions. We utilized Gemini 3 Flash \citep{google2025gemini3flash} as the backbone LLM. All details are in \cref{app:tree_search}.

\subsection{Data and Code}
\label{sec:code}

We will release datasets and code, including the JAX-translated neural circuit model, upon publication.

\section{Results}

\subsection{Sensory Drive and Identifiability}
A fundamental challenge in neural modeling is distinguishing intrinsic dynamics from responses to extrinsic drive \citep{vahidi2024modeling}. We used our \textit{in silico} testbed to perform this decomposition explicitly---an intervention virtually impossible to replicate in real-world experiments. We evaluated a model reflecting the ground truth circuit in two regimes: conditioned on instantaneous sensory drive and the history of past activity (\model{gt}[s+h]), and a control for which the sensory drive $\Phi(\mathbf{u}^{\text{ret}}_t)$ is set to zero (\model{gt}[h]).

\begin{figure*}[t]
  \begin{center}
    \centerline{\includegraphics[width=\linewidth]{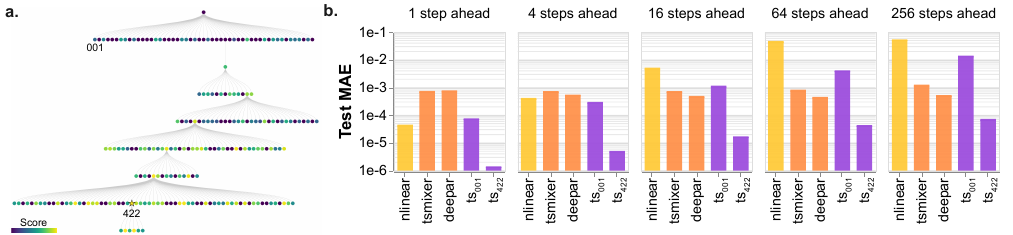}}
    \caption{
      \textbf{Tree search discovers SOTA predictive models}.
      \textbf{a.}~Progression of LLM-guided tree search, highlighting first and highest-scoring solutions (\model{ts}[001], \model{ts}[422]). For visual clarity, this tree is pruned to show only the ancestral path of \model{ts}[422]; the complete, unpruned search tree is in \cref{fig:trees}.
      \textbf{b.}~Performance comparison of discovered models against human-curated baselines. The highest-scoring discovered tree-search architecture (\model{ts}[422]) significantly outperforms baselines across all prediction horizons.
    }
    \label{fig:ts}
    \phantomsubcaption\label{fig:ts-a}
    \phantomsubcaption\label{fig:ts-b}
  \end{center}
  \capskip
\end{figure*}

Despite using the true transition function $f^*$, \model{gt}[h] yields higher error than the naive \model{mean}[h] baseline (\cref{fig:sensory}). This apparent paradox highlights the dissipative nature of the system: without sensory drive, the underlying circuit dynamics decay to rest. In contrast, the \model{mean}[h] baseline minimizes error by predicting that neural activity persists unchanged---thereby reflecting the average activity maintained by the continuous sensory drive. This shows that in a purely autoregressive setting, the true mechanistic model is practically non-identifiable based on standard error metrics. Only when conditioning on sensory drive in addition to history is the system identification task well-posed: The error for \model{gt}[s+h] drops to the numerical floor, representing the limits of floating-point precision. 

However, access to sensory drive is necessary but not sufficient; the model architecture must also be capable of effectively utilizing it. While the multivariate \model{tsmixer}[s+h] and \model{deepar}[s+h] baselines improve with the addition of sensory covariates, the univariate \model{tide}[s+h] model does not significantly improve over its history-only counterpart, \model{tide}[h]. This disparity highlights a representational mismatch: the visual stimulus is a global signal that exerts differential effects across the population. Global univariate models like \model{tide}---which apply a shared mapping across all units---lack the expressivity to capture how a single global input drives heterogeneous neural responses.

\subsection{Model Discovery with Tree Search}
\label{sec:tree_search}

The search space of predictive architectures is vast, encompassing a wide variety of structural biases, connectivity patterns, and temporal integration strategies. Traditionally, navigating this space relies on researcher intuition and manual trial-and-error. To evaluate whether automated discovery can identify accurate models of neural dynamics, we utilized an LLM-guided tree search.

In our initial experiments, tree search tended to produce heterogeneous code across various frameworks, making systematic evaluation and integration difficult. We developed a standardized evaluation harness and prompt designed specifically to search within the class of autoregressive models implemented in \texttt{flax} \citep{flax2020github}. To ensure the discovery process remained computationally efficient, we implemented a heavily optimized, GPU-resident data loading pipeline. We found that training models on a truncated rollout of 32 timesteps was sufficient to identify architectures that performed well on the full 256-step prediction horizon, significantly reducing the training time required for each candidate solution during the search process.

Following these changes, the automated discovery process \cref{fig:ts-a} was highly successful, with many of the identified solutions achieving significantly lower Test MAE than the predictive baselines: \cref{fig:ts-b} highlights the performance of two discovered models, \model{ts}[001] (the first AI-generated solution) and \model{ts}[422], compared to \texttt{nlinear}, \texttt{tsmixer}, and \texttt{deepar}. Notably, \model{ts}[422] represents the highest-scoring outcome of our search, outperforming state-of-the-art baselines by almost an order of magnitude when averaging over all prediction horizons.

\subsection{Generalization and Interpretability}
\label{sec:generalization_interpretability}

\begin{figure*}[t]
  \begin{center}
    \centerline{\includegraphics[width=\linewidth]{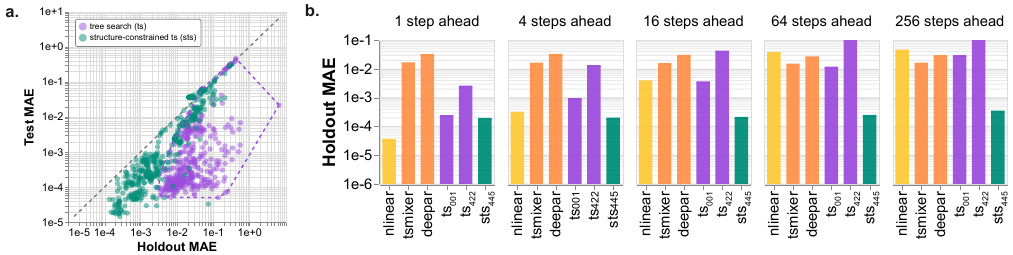}}
    \caption{
      \textbf{Structural priors enable robust generalization.}
  \textbf{a.}~Generalization analysis comparing performance on in-distribution (Test MAE) versus out-of-distribution stimuli (Holdout MAE). Tree search models (\texttt{ts}, purple) exhibit a significant generalization gap; while they achieve high accuracy on the test set, they fail to generalize to novel stimuli, suggesting they memorize sensory-motor correlations. Structurally-constrained tree search solutions (\texttt{sts}, green) cluster closer along the diagonal, indicating robust transfer to unseen environments. Dotted purple line marks the convex hull for \texttt{ts} solutions.
  \textbf{b.}~Holdout performance across prediction horizons. The highest-scoring constrained model (\model{sts}[445]) significantly outperforms its unconstrained counterpart (\model{ts}[422]) and baselines.
    }
    \label{fig:sts}
    \phantomsubcaption\label{fig:sts-a}
    \phantomsubcaption\label{fig:sts-b}
  \end{center}
  \capskip
\end{figure*}

Despite the high predictive accuracy of the tree-search models, they have limited scientific utility: they fail to generalize and it is unclear how to map them to the underlying circuit.

As shown in \cref{fig:sts-a}, unconstrained tree-search models (\texttt{ts}) with low Test MAE exhibit a significant generalization gap. Although they achieve near-perfect MAE on the test set for trained stimulus conditions, their performance degrades sharply when evaluated on held-out visual stimuli. This indicates that the search identifies solutions that exploit stimulus-specific correlations---essentially capturing statistical shortcuts---rather than recovering mechanisms that govern the zebrafish’s response.

Furthermore, these solutions offer limited interpretability for the computational neuroscientist. In \cref{app:ts_422}, we show the transition function for the highest-scoring model, \model{ts}[422]. While the code is mathematically precise, it is mechanistically opaque. The transition function lacks a clear correspondence to the neural circuit and instead makes its predictions in a 256-dimensional latent space. This lack of mechanistic transparency makes it difficult to map the model's internal operations back to the physical circuit, rendering it of limited value for formulating testable biological hypotheses.

\subsection{Incorporating Structural Priors}
\label{sec:structural_constraints}

To bridge the gap between predictive accuracy and scientific utility, we modified the discovery process by incorporating structural priors. Instead of an unconstrained search over transition functions, we provided the tree search with prior information derived from the structural connectivity of the neural circuit (\cref{app:prompts}): These hints reveal the system's hierarchical organization, the sparse connectivity patterns (including recurrence) and the functional role of covariates. Crucially, the structural prior provided to the search contains no information about connectivity weights---only about the existence and absence of connections. The specific non-linear functions that govern the interactions must still be discovered.

The impact of the structural prior is clearly evident in the model's ability to generalize. As shown in \cref{fig:sts-a}, structure-constrained models (green markers, \model{sts}) are closer to the diagonal region of the scatter plot. Unlike the unconstrained models, \model{sts} solutions maintain more consistent performance across both training and held-out stimulus conditions, suggesting that they overfit less to correlations present in train and test sets. The highest-scoring model (\model{sts}[445]) is significantly more robust on the OOD condition, more than an order of magnitude for longer prediction horizons compared to baselines (\cref{fig:sts-b}).

In Appendix~\ref{app:sts_445}, we show the transition function for \model{sts}[445]: In contrast to the hard-to-interpret code of \model{ts}[422], it reflects the structural organization of the circuit and is composed of semantically meaningful motifs.

\subsection{Mechanistic Recovery}

\begin{figure*}[t]
  \begin{center}
    \centerline{\includegraphics[width=\linewidth]{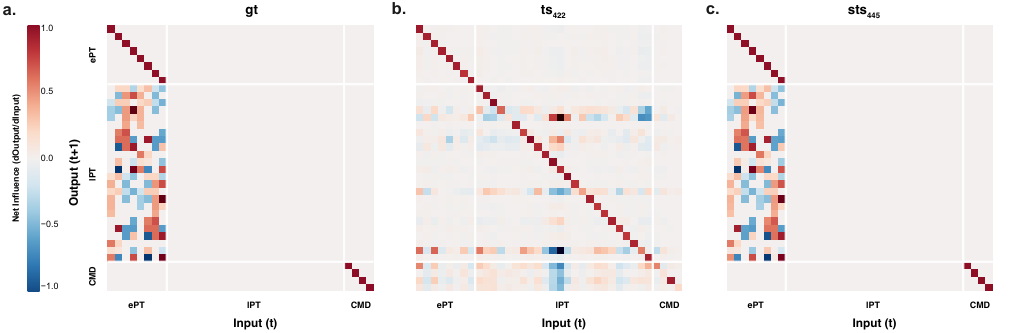}}
    \caption{\textbf{Structural priors enable mechanistic recovery.} 
    \textbf{a.} Effective connectivity matrices derived via Jacobian sensitivity analysis for the ground-truth circuit. Note the distinct excitatory (red) and inhibitory (blue) pattern in the ePT $\to$ lPT interactions (middle-left block) and the absence of recurrence in the lPT block (center).
    \textbf{b.} The unconstrained \texttt{ts}$_{422}$ model largely fails to capture the ePT $\to$ lPT interactions, instead relying on spurious recurrent lPT dynamics.
    \textbf{c.} The structure-constrained model, which received a boolean mask for ePT $\to$ lPT connectivity but no information about strength, faithfully recovers the sign and magnitude pattern of the interactions.
    }
    \label{fig:mech}
    \phantomsubcaption\label{fig:mech-a}
    \phantomsubcaption\label{fig:mech-b}
    \phantomsubcaption\label{fig:mech-c}
  \end{center}
  \capskip
\end{figure*}

We analyzed the effective connectivity of the highest-scoring discovered solution to evaluate mechanistic recovery (\cref{app:jac}). In the ground truth circuit (\cref{fig:mech-a}), the lPT population depends on structured inputs from the ePT population (middle-left block). The unconstrained \model{ts}[422] model (\cref{fig:mech-b}) fails to recover these upstream dependencies. Instead, it relies on spurious recurrence within the lPT population (central block)---effectively substituting internal memory for sensory processing.

In contrast, \model{sts}[445] recovers the correct effective connectivity.  While the structural prior included a binary matrix for ePT $\to$ lPT connectivity and indicated absence of recurrent lPT connections, it contained no information about the strengths of the feedforward interactions. Consequently, the optimization process had to discover the nature of these links. We find that \texttt{sts}$_{445}$ correctly recovered the excitatory and inhibitory checkerboard pattern of ePT $\to$ lPT interactions, matching the ground truth in both sign and relative magnitude (compare \cref{fig:mech-a} versus \cref{fig:mech-c}).

To analyze the fidelity of dynamic mechanisms of the discovered models, we performed an impulse response analysis (\cref{app:impulse_responses}), perturbing the system to a saturated state and observing the relaxation dynamics under different behavioral contexts (swimming versus resting): While \model{ts}[422] produces spurious oscillations and fails to capture the correct decay rates, \model{sts}[445] closely tracks the ground truth trajectories as shown in \cref{fig:impulse_responses}. 

We confirm the qualitative differences discussed in this section by quantitative metrics reported in \cref{tab:results_best}: We find that, relative to \model{ts}[422], \model{sts}[445] achieves a more than 30-fold reduction in effective connectivity error ($\mathcal{L}_{\text{Jac}}$), and a more than 5-fold reduction in impulse response error ($\mathcal{L}_{\text{IR}}$). \cref{tab:results_median} reports the same metrics for the top 50 highest-scoring solutions of both tree searches, confirming the robustness of these results.

Taken together, these analyses show that structural priors enable the discovery of models that abandon statistical shortcuts in favor of recovering true mechanisms, allowing faithful system identification.

\section{Discussion}
\label{sec:discussion}

Our analysis reveals that sensory drive is a prerequisite for identifiability, while in-distribution accuracy is a poor proxy for system identification. Crucially, we find that structural priors enable mechanistic recovery, eliminating statistical shortcuts that unconstrained models exploit. These insights motivate three concrete recommendations for future modeling of real-world datasets:

\textbf{\#1: Redefining prediction tasks.} Current benchmarks, such as ZAPBench \citep{lueckmann2025zapbench} which is based on whole-brain neural activity recordings from a larval zebrafish, treat neural activity as a self-contained time series suitable for autoregressive prediction. However, our control experiments show that without access to the driving sensory signals, the true circuit mechanisms may be non-identifiable. In real-world recordings, this issue is exacerbated by additional unobserved drivers (e.g., olfactory, auditory, or mechanosensory inputs). We propose changing the task from predicting future activity from history and stimulus to predicting downstream integration, conditioned on the activity of sensory populations identified via the connectome.

\textbf{\#2: OOD generalization as the primary metric}. We show that unconstrained models can achieve near-perfect accuracy on familiar data by exploiting statistical shortcuts, only to fail completely when the stimulus changes. To ensure models capture true mechanisms rather than dataset-specific correlations, evaluation should prioritize performance on held-out stimulus classes or perturbation studies.

\textbf{\#3: Use physical wiring as a scaffold.}
We propose a shift toward structural grey-box modeling. While black-box models currently dominate neural forecasting, our \textit{in silico} experiments suggest they poorly recover mechanisms and tend to have limited interpretability.

The upcoming release of the synapse-resolution connectome for the ZAPBench specimen will make these recommendations practical and offer an opportunity to test a key result from our work: that static wiring diagrams can serve as potent scaffold. 
It is often assumed that recovering mechanisms requires detailed knowledge of synaptic strengths and valences from the ultrastructure or molecular annotations.
However, we find that knowing the wiring diagram provides a structural prior that allows an optimization process to infer mechanisms directly from paired neural activity.

\textbf{Limitations and Future Work.} 
simZFish is a simplification of the vertebrate brain, focusing on a particular visuomotor circuit and associated behavior. Future work could extend the simulator to include broader behavioral repertoires and associated neural circuits. Additionally, while the tree search is applicable to high-dimensional data, the process is computationally intensive. Finally, we prioritized a fully observable, high-fidelity regime to rigorously validate mechanistic recovery against a known ground truth. Extending this to settings with partial observability or lower signal-to-noise ratios is an area for future research.

\textbf{Conclusion.} 
\textit{In silico} testbeds allow us to better understand the capabilities of AI-driven code search. Our results reveal that treating model discovery purely as an unconstrained error-minimization problem encourages solutions that exploit statistical shortcuts rather than faithful system identification. As we deploy these tools on complex real-world science problems, testbeds with verifiable solutions can guide the design of constraints and metrics needed to align AI-driven discovery with scientific utility and mechanistic understanding.


\section*{Acknowledgements}

We would like to thank Peter H. Li, Franz Rieger, Zhengdao Chen, and the Science AI team at Google Research for discussions and project support. We also thank the authors of open-source software enabling this work, including simZFish \citep{liu2024artificial}, Webots \citep{michel2004webots},  JAX \citep{jax2018github}, Flax \citep{flax2020github}, and the wider Python community.

\bibliography{main}

\begin{thebibliography}{26}
\providecommand{\natexlab}[1]{#1}
\providecommand{\url}[1]{\texttt{#1}}
\expandafter\ifx\csname urlstyle\endcsname\relax
  \providecommand{\doi}[1]{doi: #1}\else
  \providecommand{\doi}{doi: \begingroup \urlstyle{rm}\Url}\fi

\bibitem[Achterberg et~al.(2023)Achterberg, Akarca, Strouse, Duncan, and Astle]{achterberg2023spatially}
J.~Achterberg, D.~Akarca, D.~Strouse, J.~Duncan, and D.~E. Astle.
\newblock Spatially embedded recurrent neural networks reveal widespread links between structural and functional neuroscience findings.
\newblock \emph{Nature Machine Intelligence}, 2023.
\newblock URL \url{https://doi.org/10.1038/s42256-023-00748-9}.

\bibitem[Aygün et~al.(2025)Aygün, Belyaeva, Comanici, Coram, Cui, Garrison, Kast, McLean, Norgaard, Shamsi, Smalling, Thompson, Venugopalan, Williams, He, Martinson, Plomecka, Wei, Zhou, Zhu, Abraham, Brand, Bulanova, Cardille, Co, Ellsworth, Joseph, Kane, Krueger, Kartiwa, Liebling, Lueckmann, Raccuglia, Wang, Chou, Manyika, Matias, Platt, Dorfman, Mourad, and Brenner]{aygun2025ai}
E.~Aygün, A.~Belyaeva, G.~Comanici, M.~Coram, H.~Cui, J.~Garrison, R.~J.~A. Kast, C.~Y. McLean, P.~Norgaard, Z.~Shamsi, D.~Smalling, J.~Thompson, S.~Venugopalan, B.~P. Williams, C.~He, S.~Martinson, M.~Plomecka, L.~Wei, Y.~Zhou, Q.-Z. Zhu, M.~Abraham, E.~Brand, A.~Bulanova, J.~A. Cardille, C.~Co, S.~Ellsworth, G.~Joseph, M.~Kane, R.~Krueger, J.~Kartiwa, D.~Liebling, J.-M. Lueckmann, P.~Raccuglia, X.~Wang, K.~Chou, J.~Manyika, Y.~Matias, J.~C. Platt, L.~Dorfman, S.~Mourad, and M.~P. Brenner.
\newblock An {{AI}} system to help scientists write expert-level empirical software.
\newblock \emph{arXiv preprint}, 2025.
\newblock URL \url{https://doi.org/10.48550/arXiv.2509.06503}.

\bibitem[Beiran and Litwin-Kumar(2025)]{beiran2025prediction}
M.~Beiran and A.~Litwin-Kumar.
\newblock Prediction of neural activity in connectome-constrained recurrent networks.
\newblock \emph{Nature Neuroscience}, 2025.
\newblock URL \url{https://doi.org/10.1038/s41593-025-02080-4}.

\bibitem[Castro et~al.(2025)Castro, Tomasev, Anand, Sharma, Mohanta, Dev, Perlin, Jain, Levin, Elteto, Dabney, Novikov, Turner, Eckstein, Daw, Miller, and Stachenfeld]{castro2025discovering}
P.~S. Castro, N.~Tomasev, A.~Anand, N.~Sharma, R.~Mohanta, A.~Dev, K.~Perlin, S.~Jain, K.~Levin, N.~Elteto, W.~Dabney, A.~Novikov, G.~C. Turner, M.~K. Eckstein, N.~D. Daw, K.~J. Miller, and K.~Stachenfeld.
\newblock Discovering symbolic cognitive models from human and animal behavior.
\newblock In \emph{International Conference on Machine Learning}, 2025.
\newblock URL \url{https://openreview.net/forum?id=dhRXGWJ027}.

\bibitem[Chen et~al.(2023)Chen, Li, Yoder, Arik, and Pfister]{chen2023tsmixer}
S.-A. Chen, C.-L. Li, N.~Yoder, S.~O. Arik, and T.~Pfister.
\newblock {{TSMixer}}: An all-{{MLP}} architecture for time series forecasting.
\newblock \emph{Transactions on Machine Learning Research}, 2023.
\newblock URL \url{https://openreview.net/forum?id=wbpxTuXgm0}.

\bibitem[Das et~al.(2023)Das, Kong, Leach, Mathur, Sen, and Yu]{das2024longterm}
A.~Das, W.~Kong, A.~Leach, S.~Mathur, R.~Sen, and R.~Yu.
\newblock Long-term {{forecasting}} with {{TiDE}}.
\newblock \emph{Transactions on Machine Learning Research}, 2023.
\newblock URL \url{https://openreview.net/forum?id=pCbC3aQB5W}.

\bibitem[Duan et~al.(2025)Duan, Dong, and Fiete]{duan2025synapses}
S.~Duan, L.~L. Dong, and I.~Fiete.
\newblock From synapses to dynamics: Obtaining function from structure in a connectome constrained model of the head direction circuit.
\newblock \emph{bioRxiv preprint}, 2025.
\newblock URL \url{https://doi.org/10.1101/2025.05.26.655406}.

\bibitem[{Google}(2025)]{google2025gemini3flash}
{Google}.
\newblock Gemini 3 {{Flash}}: Frontier intelligence built for speed, December 17 2025.
\newblock URL \url{https://blog.google/products-and-platforms/products/gemini/gemini-3-flash/}.
\newblock The Keyword.

\bibitem[Han et~al.(2023)Han, Poggio, and Cheung]{han2023system}
Y.~Han, T.~A. Poggio, and B.~Cheung.
\newblock System identification of neural systems: If we got it right, would we know?
\newblock In \emph{International Conference on Machine Learning}, 2023.
\newblock URL \url{https://openreview.net/forum?id=NkTEhPQCjg}.

\bibitem[Jiang et~al.(2025)Jiang, Schmidt, Srikanth, Xu, Kaplan, Jacenko, and Wu]{jiang2025aide}
Z.~Jiang, D.~Schmidt, D.~Srikanth, D.~Xu, I.~Kaplan, D.~Jacenko, and Y.~Wu.
\newblock Aide: Ai-driven exploration in the space of code.
\newblock \emph{arXiv preprint}, 2025.
\newblock URL \url{https://doi.org/10.48550/arXiv.2502.13138}.

\bibitem[Jonas and Kording(2017)]{jonas2017could}
E.~Jonas and K.~P. Kording.
\newblock Could a neuroscientist understand a microprocessor?
\newblock \emph{PLoS computational biology}, 2017.
\newblock URL \url{https://doi.org/10.1371/journal.pcbi.1005268}.

\bibitem[Lange et~al.(2025)Lange, Imajuku, and Cetin]{lange2025shinkaevolve}
R.~T. Lange, Y.~Imajuku, and E.~Cetin.
\newblock {{ShinkaEvolve}}: Towards open-ended and sample-efficient program evolution.
\newblock \emph{arXiv preprint}, 2025.
\newblock URL \url{https://doi.org/10.48550/arXiv.2509.19349}.

\bibitem[Lappalainen et~al.(2024)Lappalainen, Tschopp, Prakhya, McGill, Nern, Shinomiya, Takemura, Gruntman, Macke, and Turaga]{lappalainen2024connectome}
J.~K. Lappalainen, F.~D. Tschopp, S.~Prakhya, M.~McGill, A.~Nern, K.~Shinomiya, S.-y. Takemura, E.~Gruntman, J.~H. Macke, and S.~C. Turaga.
\newblock Connectome-constrained networks predict neural activity across the fly visual system.
\newblock \emph{Nature}, 2024.
\newblock URL \url{https://doi.org/10.1038/s41586-024-07939-3}.

\bibitem[Lobato-Rios et~al.(2022)Lobato-Rios, Ramalingasetty, {\"O}zdil, Arreguit, Ijspeert, and Ramdya]{lobato2022neuromechfly}
V.~Lobato-Rios, S.~T. Ramalingasetty, P.~G. {\"O}zdil, J.~Arreguit, A.~J. Ijspeert, and P.~Ramdya.
\newblock {{NeuroMechFly}}, a neuromechanical model of adult drosophila melanogaster.
\newblock \emph{Nature Methods}, 2022.
\newblock URL \url{https://doi.org/10.1038/s41592-022-01466-7}.

\bibitem[Loshchilov and Hutter(2019)]{loshchilov2017fixing}
I.~Loshchilov and F.~Hutter.
\newblock Decoupled weight decay regularization.
\newblock In \emph{International Conference on Learning Representations}, 2019.
\newblock URL \url{https://openreview.net/forum?id=Bkg6RiCqY7}.

\bibitem[Mi et~al.(2021)Mi, Xu, Prakhya, Lin, Shavit, Samuel, and Turaga]{mi2021connectome}
L.~Mi, R.~Xu, S.~Prakhya, A.~Lin, N.~Shavit, A.~Samuel, and S.~C. Turaga.
\newblock Connectome-constrained latent variable model of whole-brain neural activity.
\newblock In \emph{International Conference on Learning Representations}, 2021.
\newblock URL \url{https://openreview.net/forum?id=CJzi3dRlJE-}.

\bibitem[Novikov et~al.(2025)Novikov, Vũ, Eisenberger, Dupont, Huang, Wagner, Shirobokov, Kozlovskii, Ruiz, Mehrabian, Kumar, See, Chaudhuri, Holland, Davies, Nowozin, Kohli, and Balog]{novikov2025alphaevolve}
A.~Novikov, N.~Vũ, M.~Eisenberger, E.~Dupont, P.-S. Huang, A.~Z. Wagner, S.~Shirobokov, B.~Kozlovskii, F.~J.~R. Ruiz, A.~Mehrabian, M.~P. Kumar, A.~See, S.~Chaudhuri, G.~Holland, A.~Davies, S.~Nowozin, P.~Kohli, and M.~Balog.
\newblock Alphaevolve: A coding agent for scientific and algorithmic discovery.
\newblock \emph{arXiv preprint}, 2025.
\newblock URL \url{https://doi.org/10.48550/arXiv.2506.13131}.

\bibitem[Romera-Paredes et~al.(2024)Romera-Paredes, Barekatain, Novikov, Balog, Kumar, Dupont, Ruiz, Ellenberg, Wang, Fawzi, Kohli, and Fawzi]{romera2024mathematical}
B.~Romera-Paredes, M.~Barekatain, A.~Novikov, M.~Balog, M.~P. Kumar, E.~Dupont, F.~J.~R. Ruiz, J.~S. Ellenberg, P.~Wang, O.~Fawzi, P.~Kohli, and A.~Fawzi.
\newblock Mathematical discoveries from program search with large language models.
\newblock \emph{Nature}, 2024.
\newblock URL \url{https://doi.org/10.1038/s41586-023-06924-6}.

\bibitem[Salinas et~al.(2020)Salinas, Flunkert, Gasthaus, and Januschowski]{salinas2020deepar}
D.~Salinas, V.~Flunkert, J.~Gasthaus, and T.~Januschowski.
\newblock Deepar: Probabilistic forecasting with autoregressive recurrent networks.
\newblock \emph{International Journal of Forecasting}, 2020.
\newblock URL \url{https://doi.org/10.1016/j.ijforecast.2019.07.001}.

\bibitem[Sarma et~al.(2018)Sarma, Lee, Portegys, Ghayoomie, Jacobs, Alicea, Cantarelli, Currie, Gerkin, Gingell, Gleeson, Gordon, Hasani, Idili, Khayrulin, Lung, Palyanov, Watts, and Larson]{sarma2018openworm}
G.~P. Sarma, C.~W. Lee, T.~Portegys, V.~Ghayoomie, T.~Jacobs, B.~Alicea, M.~Cantarelli, M.~Currie, R.~C. Gerkin, S.~Gingell, P.~Gleeson, R.~Gordon, R.~M. Hasani, G.~Idili, S.~Khayrulin, D.~Lung, A.~Palyanov, M.~Watts, and S.~D. Larson.
\newblock {{OpenWorm}}: Overview and recent advances in integrative biological simulation of caenorhabditis elegans.
\newblock \emph{Philosophical Transactions of the Royal Society B}, 2018.
\newblock URL \url{https://doi.org/10.1098/rstb.2017.0382}.

\bibitem[Schmidt and Lipson(2009)]{schmidt2009distilling}
M.~Schmidt and H.~Lipson.
\newblock Distilling free-form natural laws from experimental data.
\newblock \emph{Science}, 324, 2009.
\newblock URL \url{https://doi.org/10.1126/science.1165893}.

\bibitem[Tilbury et~al.(2025)Tilbury, Kwon, Haydaroglu, Ratliff, Schmutz, Carandini, Miller, Stachenfeld, and Harris]{tilbury2025characterizing}
R.~Tilbury, D.~Kwon, A.~Haydaroglu, J.~Ratliff, V.~Schmutz, M.~Carandini, K.~Miller, K.~Stachenfeld, and K.~D. Harris.
\newblock Characterizing neuronal population geometry with {{AI}} equation discovery.
\newblock \emph{bioRxiv preprint}, 2025.
\newblock URL \url{https://doi.org/10.1101/2025.11.12.688086}.

\bibitem[Vaxenburg et~al.(2025)Vaxenburg, Siwanowicz, Merel, Robie, Morrow, Novati, Stefanidi, Both, Card, Reiser, Botvinick, Branson, Tassa, and Turaga]{vaxenburg2025whole}
R.~Vaxenburg, I.~Siwanowicz, J.~Merel, A.~A. Robie, C.~Morrow, G.~Novati, Z.~Stefanidi, G.-J. Both, G.~M. Card, M.~B. Reiser, M.~M. Botvinick, K.~M. Branson, Y.~Tassa, and S.~C. Turaga.
\newblock Whole-body physics simulation of fruit fly locomotion.
\newblock \emph{Nature}, 2025.
\newblock URL \url{https://doi.org/10.1038/s41586-025-09029-4}.

\bibitem[Zeng et~al.(2023)Zeng, Chen, Zhang, and Xu]{zeng2022are}
A.~Zeng, M.~Chen, L.~Zhang, and Q.~Xu.
\newblock Are {{transformers effective}} for {{time series forecasting}}?
\newblock In \emph{AAAI Conference on Artificial Intelligence}, 2023.
\newblock URL \url{https://doi.org/10.1609/aaai.v37i9.26317}.

\bibitem[Zhao et~al.(2024)Zhao, Wang, Jiang, Ma, Ma, He, Du, Ma, and Huang]{zhao2025integrative}
M.~Zhao, N.~Wang, X.~Jiang, X.~Ma, H.~Ma, G.~He, K.~Du, L.~Ma, and T.~Huang.
\newblock An integrative data-driven model simulating c. elegans brain, body and environment interactions.
\newblock \emph{Nature Computational Science}, 2024.
\newblock URL \url{https://doi.org/10.1038/s43588-024-00738-w}.

\bibitem[Zoph and Le(2017)]{zoph2017neural}
B.~Zoph and Q.~V. Le.
\newblock Neural architecture search with reinforcement learning.
\newblock In \emph{International Conference on Learning Representations}, 2017.
\newblock URL \url{https://openreview.net/forum?id=r1Ue8Hcxg}.

\end{thebibliography}

\newpage
\appendix
\onecolumn

\crefalias{section}{appendix}
\crefalias{subsection}{subappendix}

\let\addcontentsline\standardaddcontentsline

\appendixpage
\startcontents[section]
\printcontents[section]{l}{0}{\setcounter{tocdepth}{3}}
\setcounter{page}{1}
\setcounter{section}{0}

\setcounter{equation}{0}
\setcounter{table}{0}
\setcounter{figure}{0}

\renewcommand\theequation{S\arabic{equation}}
\renewcommand\thetable{S\arabic{table}}
\renewcommand\thefigure{S\arabic{figure}}

\newpage

\section{Related Work}
\label{app:related_work}

\textbf{Automated Model Discovery.} 
Our work uses the capability of Large Language Models (LLMs) to perform automated model discovery via code generation. While traditional approaches to model discovery relied on genetic programming \citepapp{schmidt2009distilling} or reinforcement learning for neural architecture search \citepapp{zoph2017neural}, recent methods treat the search process as an evolution of executable code. We utilize the tree-search methodology proposed by \citetapp{aygun2025ai}. Parallel efforts have explored similar LLM-driven evolutionary strategies for mathematical discovery \citepapp{romera2024mathematical} and algorithmic evolution \citepapp{novikov2025alphaevolve,lange2025shinkaevolve,jiang2025aide}. In related domains, automated model discovery techniques have for example been applied to search for interpretable cognitive models of human and animal behavior \citepapp{castro2025discovering}, and neural tuning curves \citepapp{tilbury2025characterizing}. Our work distinguishes itself by applying these methods to neural system identification in a setting where we can rigorously benchmark the discovery of mechanisms against a verifiable ground truth.

\textbf{System Identification with Ground Truth.} 
The difficulty of inferring ground-truth mechanisms from neural data is a long-standing methodological concern. \citetapp{jonas2017could} utilized a microprocessor to demonstrate that standard data analysis techniques can fail to recover the known logic of information processing. More recently, \citetapp{han2023system} used artificial neural networks (ANNs) as in silico ground truths, revealing that functional similarity (predictive performance) does not imply mechanistic similarity. We advance this line of inquiry by moving beyond artificial proxies---such as microprocessors or generic ANNs---which differ fundamentally from biological substrates. Instead, we employ a high-fidelity, experimentally constrained emulation of the zebrafish optomotor response to move towards more realistic models with ground truth.

\textbf{Neuromechanical Simulations and Connectome Constraints.} 
Our work is situated within the growing field of neuromechanical modeling. While detailed simulations have been established for invertebrates such as the worm \textit{C. elegans} \citepapp{sarma2018openworm, zhao2025integrative} and the fruit fly \textit{D. melanogaster} \citepapp{lobato2022neuromechfly,vaxenburg2025whole}, we utilize a larval zebrafish simulation to capture vertebrate visuomotor dynamics. Concurrently, the increasing availability of connectomes has spurred research into constraining neural models with physical wiring diagrams. The success of our structurally constrained models aligns with the findings of \citet{sourmpis2025biologically}, who demonstrated that biologically informed models are more robust to optogenetic perturbations than generic RNNs. Several studies, including \citetapp{mi2021connectome,achterberg2023spatially,lappalainen2024connectome,beiran2025prediction,duan2025synapses}, have highlighted the predictive utility of structural constraints. Our work bridges these findings with automated discovery: rather than manually designing constrained architectures, we show that connectomic information can be used with AI-driven search to navigate the vast space of possible models and recover mechanistic solutions.

\newpage

\section{Neuromechanical Simulations}
\label{app:simulations}

We detail the numerical values for the parameters introduced in the neural circuit model (\cref{sec:neuromechanical_simulations}), which are based on the reference implementation \cite{liu2024artificial}.

\textbf{Early Pretectum (ePT)}. The ePT dynamics described in \cref{eq:ept_layer} are governed by the update rate $\alpha_t$ and the response function parameters $\boldsymbol{\omega}_{\text{ept}}, \mathbf{b}_{\text{ept}}$:
\begin{align*}
\alpha_t &= 
\begin{cases} 
0.002 & \text{if swim bout active,} \\
0.01 & \text{otherwise,}
\end{cases} \\
\boldsymbol{\omega}_{\text{ept}} &= [0.068, 0.100, 0.068, 0.100, 0.068, 0.100, 0.068, 0.100]^\top, \\
\mathbf{b}_{\text{ept}} &= [90.0, 60.0, 90.0, 60.0, 90.0, 60.0, 90.0, 60.0]^\top.
\end{align*}

The ePT state vector is ordered as $\mathbf{h}^{\text{ept}} = [\text{ePT}_\text{L}^\text{S}, \text{ePT}_\text{L}^\text{A}, \text{ePT}_\text{L}^\text{I}, \text{ePT}_\text{L}^\text{P}, \text{ePT}_\text{R}^\text{S}, \text{ePT}_\text{R}^\text{A}, \text{ePT}_\text{R}^\text{I}, \text{ePT}_\text{R}^\text{P}]^\top$, i.e., grouped by hemisphere (Left, Right) and functional tuning (Superior, Anterior, Inferior, Posterior).

\textbf{Late Pretectum (lPT)}. The lPT activations in \cref{eq:lpt_layer} are computed using the connectivity matrix $\mathbf{W}_{\text{lpt}}$, the gating mask $\mathbf{M}_{\text{gate}}$, and response parameters $\boldsymbol{\omega}_{\text{lpt}}, \mathbf{b}_{\text{lpt}} \in \mathbb{R}^{24}$:
\begin{align*}
\mathbf{W}_{\text{lpt}} &= \left[
\begin{smallmatrix}
0 & 1 & 1 & -1.5 & 0.8 & 1.5 & -1 & -2 \\
0 & -1 & 1 & 1.5 & 0.8 & 0 & -1 & 0 \\
-0.8 & 1 & 1 & 0 & 0.8 & 1 & -1 & -2 \\
-0.8 & -1 & 1 & 2.5 & 0.8 & 0 & -1 & 0 \\
-0.8 & 0 & 1 & 0.5 & 0.8 & 0 & 0 & 0 \\
0 & 0 & 1 & 0 & 0.8 & 0 & 0 & 0 \\
0 & 0.8 & 1 & 0 & 0 & 0 & 0 & -0.8 \\
0 & 1 & 1 & -1 & 0 & 1.5 & -1 & -1 \\
0 & -1.5 & 1 & 1.5 & 0 & 0 & -1 & 0.5 \\
0 & 0 & 0 & 0 & 0.8 & 0.3 & 0 & 0 \\
0 & 1.5 & 1 & -1 & 0.8 & 0.5 & 0 & -2 \\
0 & -1.5 & 1 & 1.5 & 0.8 & -1 & 0 & 1 \\
0.8 & 1.5 & -1 & -2 & 0 & 1 & 1 & -1.5 \\
0.8 & 0 & -1 & 0 & 0 & -1 & 1 & 1.5 \\
0.8 & 1 & -1 & -2 & -0.8 & 1 & 1 & 0 \\
0.8 & 0 & -1 & 0 & -0.8 & -1 & 1 & 2.5 \\
0.8 & 0 & 0 & 0 & -0.8 & 0 & 1 & 0.5 \\
0.8 & 0 & 0 & 0 & 0 & 0 & 1 & 0 \\
0 & 0 & 0 & -0.8 & 0 & 0.8 & 1 & 0 \\
0 & 1.5 & -1 & -1 & 0 & 1 & 1 & -1 \\
0 & 0 & -1 & 0.5 & 0 & -1.5 & 1 & 1.5 \\
0.8 & 0.3 & 0 & 0 & 0 & 0 & 0 & 0 \\
0.8 & 0.5 & 0 & -2 & 0 & 1.5 & 1 & -1 \\
0.8 & -1 & 0 & 1 & 0 & -1.5 & 1 & 1.5
\end{smallmatrix}
\right], \quad
\mathbf{M}_{\text{gate}} = \left[
\begin{smallmatrix}
0 \\ 0 \\ 0 \\ 0 \\ 0 \\ 0 \\ 0 \\ 0 \\ 0 \\ 0 \\ 1 \\ 1 \\
0 \\ 0 \\ 0 \\ 0 \\ 0 \\ 0 \\ 0 \\ 0 \\ 0 \\ 0 \\ 1 \\ 1
\end{smallmatrix}
\right], \quad
\boldsymbol{\omega}_{\text{lpt}} = \left[
\begin{smallmatrix}
2.5 \\ 2.5 \\ 2.5 \\ 2.5 \\ 2.5 \\ 2.5 \\ 4.0 \\ 4.0 \\ 4.0 \\ 5.0 \\ 4.0 \\ 4.0 \\
2.5 \\ 2.5 \\ 2.5 \\ 2.5 \\ 2.5 \\ 2.5 \\ 4.0 \\ 4.0 \\ 4.0 \\ 5.0 \\ 4.0 \\ 4.0
\end{smallmatrix}
\right], \quad
\mathbf{b}_{\text{lpt}} = \left[
\begin{smallmatrix}
0.8 \\ 0.8 \\ 0.8 \\ 0.8 \\ 0.8 \\ 0.8 \\ 0.6 \\ 0.6 \\ 0.6 \\ 0.4 \\ 0.6 \\ 0.6 \\
0.8 \\ 0.8 \\ 0.8 \\ 0.8 \\ 0.8 \\ 0.8 \\ 0.6 \\ 0.6 \\ 0.6 \\ 0.4 \\ 0.6 \\ 0.6
\end{smallmatrix}
\right].
\end{align*}

The lPT state vector is ordered as
\begin{equation*}
\begin{split}
\mathbf{h}^{\text{lpt}} = [ & \text{lPT}_\text{L}^{\text{oB}_1}, \text{lPT}_\text{L}^{\text{oB}_2}, \text{lPT}_\text{L}^{\text{B}_1}, \text{lPT}_\text{L}^{\text{B}_2}, \text{lPT}_\text{L}^{\text{iB}}, \text{lPT}_\text{L}^{\text{ioB}}, 
\text{lPT}_\text{L}^{\text{iMm}}, \text{lPT}_\text{L}^{\text{Mm}_1}, \text{lPT}_\text{L}^{\text{Mm}_2}, \text{lPT}_\text{L}^{\text{oMm}}, \text{lPT}_\text{L}^{\text{S}_1}, \text{lPT}_\text{L}^{\text{S}_2}, \\
& \text{lPT}_\text{R}^{\text{oB}_1}, \dots, \text{lPT}_\text{R}^{\text{S}_2}]^\top,
\end{split}
\end{equation*}
grouping first by hemisphere (Left, Right) and then by functional type.

The gating function $\mathbf{g}(\mathbf{h})$ referenced in \cref{eq:lpt_layer} implements a ``soft'' AND-gate using a minimum operation between specific pairs of inputs. This function selects specific ipsilateral ($\mathbf{h}^{\text{ipsi}}$) and contralateral ($\mathbf{h}^{\text{contra}}$) components from the ePT state vector $\mathbf{h}^{\text{ept}}$. Specifically, for ``S-type''-LPT neurons, the gating signal requires activation from both contralateral Superior and ipsilateral Inferior inputs:
\begin{equation*}
\mathbf{g}(\mathbf{h}) = \min(0.8 \cdot \mathbf{h}^{\text{contra}}, \mathbf{h}^{\text{ipsi}}) = \min(0.8 \cdot \text{ePT}_\text{R}^\text{S}, \text{ePT}_\text{L}^\text{I}).
\end{equation*}
This ensures the neuron is only active when both specific directional conditions are met, scaled by a factor of 0.8 on the contralateral input to account for asymmetric drive.

\textbf{Command Neurons (nMLF/aHB)}. The dynamics of the command neurons in \cref{eq:command_layer} are defined by the update rate $\beta = 0.0005$, connectivity $\mathbf{W}_{\text{cmd}}$, and response parameters $\boldsymbol{\omega}_{\text{cmd}}, \mathbf{b}_{\text{cmd}}$: 
\begin{align*}
\mathbf{W}_{\text{cmd}}^{1\dots12} &= \left[
\begin{smallmatrix}
0.1 & -0.1 & 0.32 & -0.08 & 0.25 & -0.05 & 0.6 & 0.4 & 0 & 0 & 0.3 & -0.1 \\
0 & 0 & 0 & 0 & 0 & 0 & 0 & 0 & 0 & -0.8 & 0 & 0 \\
0.1875 & 0.3875 & 0.2165 & 0.3165 & 0 & 0 & 0 & 0.0165 & 0.4165 & 0 & 0.096 & 0.096 \\
-0.1665 & -0.1665 & -0.1665 & -0.1665 & 0 & 0 & 0 & -0.105 & -0.105 & 0 & -0.29 & -0.29
\end{smallmatrix}
\right], \\
\mathbf{W}_{\text{cmd}}^{13\dots24} &= \left[
\begin{smallmatrix}
0 & 0 & 0 & 0 & 0 & 0 & 0 & 0 & 0 & -0.8 & 0 & 0 \\
0.1 & -0.1 & 0.32 & -0.08 & 0.25 & -0.05 & 0.6 & 0.4 & 0 & 0 & 0.3 & -0.1 \\
-0.1665 & -0.1665 & -0.1665 & -0.1665 & 0 & 0 & 0 & -0.105 & -0.105 & 0 & -0.29 & -0.29 \\
0.1875 & 0.3875 & 0.2165 & 0.3165 & 0 & 0 & 0 & 0.0165 & 0.4165 & 0 & 0.096 & 0.096
\end{smallmatrix}
\right], \\
\boldsymbol{\omega}_{\text{cmd}} &= [6.5, 6.5, 4.0, 4.0]^\top, \quad
\mathbf{b}_{\text{cmd}} = [0.4, 0.4, 0.65, 0.65]^\top.
\end{align*}
The resulting command state vector is ordered as $\mathbf{h}^{\text{cmd}}_t = [\text{nMLF}_\text{L}, \text{nMLF}_\text{R}, \text{aHB}_\text{L}, \text{aHB}_\text{R}]^\top$.

\textbf{Motor circuits.} The evidence accumulation dynamics in \cref{eq:bout_accumulation} are defined by the integrator threshold $\theta = 436.0$, noise standard deviation $\sigma = 0.65$, and nMLF input scaling factor $\gamma = 0.6$.

The action probabilities defined in \cref{eq:action_selection} are computed using the command state vector $\mathbf{h}^{\text{cmd}}_t$, the projection matrix $\mathbf{W}_{\text{act}}$, and $\mathbf{b}_{\text{act}}$:
\begin{align*}
\mathbf{W}_{\text{act}} &= \begin{bmatrix}
1 & 1 & 0 & 0 \\
0 & 0 & 1 & 0 \\
0 & 0 & 0 & 1
\end{bmatrix}, \quad
\mathbf{b}_{\text{act}} = \begin{bmatrix}
0.200 \\
0.066 \\
0.066
\end{bmatrix}.
\end{align*}

For an explanation of all other stages in the model, see \citet{liu2024artificial}.

\newpage

\section{Additional Analyses}
\label{app:additional_analyses}

\subsection{Effective Connectivity Analysis}
\label{app:jac}

To evaluate whether the discovered models recover the true static mechanisms of the neural circuit, we computed the \textit{Effective Connectivity Matrix}. This matrix quantifies the sensitivity of the predicted neural state at time $t+1$ to perturbations in the state at time $t$ (\cref{fig:mech}).

Formally, we analyze the one-step transition $\mathbf{a}_{t+1} = f_\theta(\mathbf{a}_{t}, \mathbf{x}_{t+1})$. The effective connectivity matrix $J \in \mathbb{R}^{N \times N}$ is defined as the Jacobian of the future state with respect to the current state:
\begin{equation}
    J_{ij} = \frac{\partial (\mathbf{a}_{t+1})_i}{\partial (\mathbf{a}_{t})_j}
\end{equation}
where $J_{ij}$ represents the causal influence of source neuron $j$ on target neuron $i$.

To isolate the intrinsic connectivity of the circuit, we compute $J$ while setting exogenous covariates to zero ($\mathbf{x}_{t+1} = \mathbf{0}$). To account for non-linearities (e.g., sigmoids) where gradients vanish in saturated regions, we do not evaluate $J$ at a single operating point. Instead, we estimate the expected sensitivity by averaging over $K=11$ homogeneous probe states. We define a set of scalars $\mathcal{S} = \{0.0, 0.1, \dots, 1.0\}$ and construct input vectors $\mathbf{a}^{(s)} = s \cdot \mathbf{1}$, where $\mathbf{1}$ is the all-ones vector. The reported connectivity is:
\begin{equation}
    \bar{J} = \frac{1}{|\mathcal{S}|} \sum_{s \in \mathcal{S}} J|_{\mathbf{a}_{t} = \mathbf{a}^{(s)}}
\end{equation}
We quantify the mechanistic fidelity of a model by comparing its connectivity $\bar{J}_{\text{model}}$ to the ground truth $\bar{J}_{\text{gt}}$ using the Frobenius norm:
\begin{equation}
    \mathcal{L}_{\text{jac}} = || \bar{J}_{\text{model}} - \bar{J}_{\text{gt}} ||_F
\end{equation}

\subsection{Impulse Response Analysis}
\label{app:impulse_responses}

To assess recovery of dynamic mechanisms, we performed an \textit{Impulse Response Analysis}. We probe how the system relaxes from perturbed states back to rest, and how this dynamic is modulated by behavioral states.

We simulate the system for a horizon of $H=256$ steps starting from a set of impulse conditions. We test $N+1$ initial conditions: 36 cases where a single neuron is activated ($\mathbf{a}_0 = \mathbf{e}_i$) and one case where the entire system is saturated ($\mathbf{a}_0 = \mathbf{1}$). We evolve these states using two fixed covariates $\mathbf{x}_{\text{mode}}$ to test mechanisms depending on swim state (e.g., the ePT update in \cref{eq:ept_layer}):
\begin{enumerate}
    \item \textbf{Rest:} Bout state and sensory drive set to 0.
    \item \textbf{Swim:} Bout state set to 1, sensory drive set to 0.
\end{enumerate}
We quantify the divergence between the model's predicted response traces ($\hat{\mathbf{A}}$) and the ground truth traces ($\mathbf{A}^*$) using the Mean Absolute Error (MAE) averaged over all time steps, units, and initial conditions. We report this metric separately for the two regimes:
\begin{equation}
    \mathcal{L}_{\text{IR}}^{\text{rest}} = \text{MAE}(\hat{\mathbf{A}}^{\text{rest}}, \mathbf{A}^{*\text{rest}}), \quad
    \mathcal{L}_{\text{IR}}^{\text{swim}} = \text{MAE}(\hat{\mathbf{A}}^{\text{swim}}, \mathbf{A}^{*\text{swim}})
\end{equation}

\newpage

\section{Baseline Models}
\label{app:baselines}

\subsection{\model{mean}}

The naive \model{mean} baseline repeats the per-neuron activity from the last timestep across the entire prediction horizon.

\subsection{\model{nlinear}}

For \model{nlinear} \citepapp{zeng2022are}, we use a model with context window size of 2. For optimization, we use AdamW \citepapp{loshchilov2017fixing} with a learning rate of $10^{-4}$ and weight decay of $10^{-4}$.

\subsection{\model{tide}}

For \model{tide} \citepapp{das2024longterm}, we use a model with a context window size of 2, 2 encoder and decoder layers each, a hidden layer size of 128, a decoder output dimensionality of 32, and reversible instance norm. For optimization, we use AdamW \citepapp{loshchilov2017fixing} with a learning rate of $10^{-3}$ and weight decay of $10^{-4}$.

\subsection{\model{tsmixer}}

For \model{tsmixer} \citepapp{chen2023tsmixer}, we use a model with a context window size of 2, 2 blocks, and an MLP dimension of 256, and no instance norm. For optimization, we use AdamW \citepapp{loshchilov2017fixing} with a learning rate of $10^{-3}$ and weight decay of $10^{-4}$.

\subsection{\model{deepar}}

For \model{deepar} \citepapp{salinas2020deepar}, we use a model with 2 stacked GRU layers, with a hidden size of 128. For optimization, we use AdamW \citepapp{loshchilov2017fixing} with a learning rate of $10^{-3}$ and weight decay of $10^{-4}$.

\newpage

\section{Tree Search}
\label{app:tree_search}

We formulate the model discovery process as a search problem where the state space consists of executable Python code defining neural transition functions, and the objective is to maximize predictive accuracy on the validation set.

\subsection{Search Algorithm}

We utilize the Predictor Upper Confidence Bound applied to Trees (PUCT) algorithm \citep{silver2016mastering, aygun2025ai} to balance exploration of novel solutions and exploitation (refining high-performing solutions). Unlike standard Monte Carlo Tree Search, where the algorithm traverses from the root to a leaf node for expansion, the space of possible code edits here is effectively infinite. Consequently, the algorithm treats every existing node in the tree as a candidate for expansion.

At each step, a node $u^*$ is sampled from the entire tree $\mathcal{T}$ by maximizing the PUCT objective:
\begin{equation}
    u^* = \operatorname*{argmax}_{u \in \mathcal{T}} \left( \tilde{R}(u) + c_{\text{puct}} |\mathcal{T}|^{-1} \frac{\sqrt{\sum_{v \in \mathcal{T}} V(v)}}{1 + V(u)} \right),
\end{equation}
where $V(u)$ is the visit count of node $u$, $c_{\text{puct}}$ is an exploration constant, and $|\mathcal{T}|^{-1}$ acts as a flat prior. 

The term $\tilde{R}(u)$ represents the normalized rank score of a solution. Solutions are scored by calculating the negative Mean Absolute Error (MAE) on the validation set, averaged over in-distribution conditions. These raw scores are converted into ranks $\operatorname{Rank}_{\mathcal{T}}(u)$ in ascending order (where the best solution with the least error has the highest rank), and normalized as:
\begin{equation}
    \tilde{R}(u) = \frac{\operatorname{Rank}_{\mathcal{T}}(u) - 1}{|\mathcal{T}| - 1}.
\end{equation}

Upon selecting a node $u^*$, we prompt the LLM to generate a child node by modifying the source code of $u^*$. The LLM acts as a mutation operator, guided by instructions defined in \cref{app:prompts}, and the initial code specified in \cref{app:ts_initial}. The generated code is immediately executed in a sandboxed environment to compute the validation score on the task defined in the \cref{sec:task_def}. Finally, the visit count $V$ is incremented for the selected node and backpropagated to its ancestors.

\subsection{Implementation Details}
\label{app:ts_imlementation_details}

We utilized Gemini 3 Flash \citepapp{google2025gemini3flash} as the backbone LLM for code generation and mutation. Each tree search was limited to a maximum of 500 nodes. To ensure computational feasibility on a single NVIDIA T4 GPU, we enforced a runtime limit of 12 hours per candidate solution. To further accelerate the discovery process, we employed a truncated training regime: models were trained on a short rollout of 32 timesteps. We found this efficient training proxy sufficient to identify architectures that generalize well when rolled out to the full prediction horizon (256 timesteps) used for scoring and evaluation.

\subsection{Prompts}
\label{app:prompts}

The initial prompt for tree search, without prior information about structure:

\begin{promptbox}
# Predicting neural activity

## Challenge

Your task is to build a forecasting model that **predicts future neural activity traces given past activity and covariates**.

**Data Specifications:**
- **Inputs & Outputs:** All neural activity and continuous covariates are normalized to the `[0, 1]` range.
- **Covariates (`cov_t`):** Shape `(Batch, CovFeatures=9)`.
    - `cov_t[..., 0]`: **Bout State**. A binary flag (0 or 1). This acts as a switch for the system's dynamics.
    - `cov_t[..., 1:]`: **Visual Input**. Continuous signals `[0, 1]` driving the system.

**Key Difficulties:**
1. **Autoregressive Stability:** You will develop a transition model $f(y_{t-1}, c_t) \rightarrow y_t$. This is rolled out for 32 timesteps. Small errors at step 1 can explode by step 32.
2. **OOD Generalization:** The test set contains visual patterns **not seen during training**. Your model must capture the underlying causal rules, not memorize training sequences.

## Evaluation

Performance is measured by Mean Absolute Error (MAE) on held-out trajectories. Lower is better.

## Rules

1.  **Structure:** Keep the sections `CONSTANTS`, `MODEL`, and `OPTIMIZER` in this order.
2.  **Interface - Hyperparameters:** You **must** define the following constants in the `CONSTANTS` section to control the training harness:
    *   `BATCH_SIZE`: (int) Default ~512.
    *   `MAX_STEPS`: (int) Default ~500,000.
    *   `SCHEDULED_SAMPLING_START_PROB` & `SCHEDULED_SAMPLING_FINAL_PROB`: (float) Controls teacher forcing. High start (1.0) helps convergence; low final (<0.1) enforces stability.
    *   `SCHEDULED_SAMPLING_DECAY_STEPS`: (int) Steps to decay the probability.
    *   `EARLY_STOP_PATIENCE`: (int) Steps to wait before stopping.
    *   `MIN_DELTA`: (float) Improvement threshold for early stopping.
3.  **Interface - Model:**
    - Use `flax.linen`. Class name must be `Candidate`.
    - `__call__` signature: `(self, past_pred, cov_t, train=False)`.
    - **Shape Contract:** Input `past_pred` is (B, 36). Output must be (B, 36).
4.  **Interface - Optimizer:**
    - Use `optax`. Instance must be named `optimizer`.
5.  **Constraints:** Do not modify code in `NOTE` comments.

## Strategic Approaches & Hints

To win this competition, you should explore distinct architectural ideas. Do not limit yourself to simple MLPs.
\end{promptbox}

For the tree search with structural prior, the \textbf{Strategic Approaches \& Hints} section in the prompt was extended as follows:

\begin{promptbox}
We have prior knowledge about the neuron hierarchy that is likely critical for a low score. The 36 neurons are split into three groups with specific dependencies:
1.  **EPT Neurons (Indices 0-8):** Depend on their past state (`past_pred[..., 0:8]`) and current covariates (`cov_t`). The bout state `cov_t[..., 0]` acts as a switch, modulating the neurons' update dynamics based on whether a bout is active. Each neuron i in this group (indices 0-7) is driven by a specific visual input: `cov_t[..., 1+i]` (e.g., neuron 0 depends on `cov_t[..., 1]`, neuron 1 on `cov_t[..., 2]`, and so on).
2.  **LPT Neurons (Indices 8:32):** Depend only on the **newly computed EPT activations** (current step's neurons 0-7), not on their own past activations from past_pred. The `CONNECTIVITY` variable below indicates sparse connections: `CONNECTIVITY[i][j]` is true if the i-th neuron within the LPT group depends on the j-th neuron within the EPT group. Since i ranges from 0-23 and j from 0-7, `CONNECTIVITY[0][j]` refers to dependencies of neuron 8 (first LPT neuron), `CONNECTIVITY[1][j]` refers to dependencies of neuron 9 (second LPT neuron), and so on.
3.  **aHB/nMLF Neurons (Indices 32:36):** Depend on their past state (`past_pred[..., 32:36]`) and the **newly computed** LPT activations (current step).

\end{promptbox}

... and the following connectivity matrix was included:

\begin{pythonbox}
# NOTE: Connectivity from EPT (cols) to LPT (rows)
CONNECTIVITY = [
 [False, True, True, True, True, True, True, True],
 [False, True, True, True, True, False, True, False],
 [True, True, True, False, True, True, True, True],
 [True, True, True, True, True, False, True, False],
 [True, False, True, True, True, False, False, False],
 [False, False, True, False, True, False, False, False],
 [False, True, True, False, False, False, False, True],
 [False, True, True, True, False, True, True, True],
 [False, True, True, True, False, False, True, True],
 [False, False, False, False, True, True, False, False],
 [False, True, True, True, True, True, False, True],
 [False, True, True, True, True, True, False, True],
 [True, True, True, True, False, True, True, True],
 [True, False, True, False, False, True, True, True],
 [True, True, True, True, True, True, True, False],
 [True, False, True, False, True, True, True, True],
 [True, False, False, False, True, False, True, True],
 [True, False, False, False, False, False, True, False],
 [False, False, False, True, False, True, True, False],
 [False, True, True, True, False, True, True, True],
 [False, False, True, True, False, True, True, True],
 [True, True, False, False, False, False, False, False],
 [True, True, False, True, False, True, True, True],
 [True, True, False, True, False, True, True, True]]
\end{pythonbox}

\newpage

\section{Code Listings}
\label{app:code}

\subsection{Initial Code for Tree Search}
\label{app:ts_initial}

The initial code, which the tree search was tasked to modify:

\begin{pythonbox}
#
# CONSTANTS
#

BATCH_SIZE = 512
MAX_STEPS = 500_000
SCHEDULED_SAMPLING_START_PROB = 1.0
SCHEDULED_SAMPLING_FINAL_PROB = 0.01
SCHEDULED_SAMPLING_DECAY_STEPS = 100_000
EARLY_STOP_PATIENCE = 20
MIN_DELTA = 1e-6
LEARNING_RATE = 1e-3


#
# MODEL
#

class Candidate(nn.Module):
  """Candidate transition model for neural activity."""

  @nn.compact
  def __call__(self, past_pred, cov_t, train=False):
    """
    Args:
      past_pred: (Batch, 36) - Previous neural state [0, 1].
      cov_t: (Batch, 9) - Index 0 is Bout {0,1}, 1-8 are Visual [0,1].
      train: bool - Training mode flag.
    Returns:
      (Batch, 36) - Predicted next state.
    """
    bout_state = cov_t[:, 0:1]  # Shape (B, 1)
    visual_in = cov_t[:, 1:]    # Shape (B, 8)

    # TODO: Implement your architecture here.

    # Identity baseline: predicts no change.
    return past_pred


transition_model = Candidate()  # NOTE: Do keep this line as-is.


#
# OPTIMIZER
#

optimizer = optax.adamw(learning_rate=LEARNING_RATE, weight_decay=1e-4)
\end{pythonbox}

\subsection{Solution \model{ts}[001]}
\label{app:ts_001}

\begin{pythonbox}
import jax.numpy as jnp
import flax.linen as nn
import optax

#
# CONSTANTS
#

BATCH_SIZE = 512
MAX_STEPS = 500_000
# High start prob uses ground truth; low final prob enforces autoregressive stability.
SCHEDULED_SAMPLING_START_PROB = 1.0
SCHEDULED_SAMPLING_FINAL_PROB = 0.02
SCHEDULED_SAMPLING_DECAY_STEPS = 150_000
EARLY_STOP_PATIENCE = 40
MIN_DELTA = 1e-6
LEARNING_RATE = 1e-3


#
# MODEL
#

class Candidate(nn.Module):
  """Residual transition model for neural activity."""

  @nn.compact
  def __call__(self, past_pred, cov_t, train=False):
    """
    Args:
      past_pred: (Batch, 36) - Previous neural state.
      cov_t: (Batch, 9) - Bout switch and Visual inputs.
      train: bool - Training mode.
    Returns:
      (Batch, 36) - Predicted next state.
    """
    # Combine previous state and external drivers
    # Shape: (Batch, 36 + 9) = (Batch, 45)
    x = jnp.concatenate([past_pred, cov_t], axis=-1)

    # MLP for state transition dynamics
    h = nn.Dense(512)(x)
    h = nn.LayerNorm()(h)
    h = nn.silu(h)

    h = nn.Dense(512)(h)
    h = nn.LayerNorm()(h)
    h = nn.silu(h)

    # Predict the delta (residual). Initialize with zeros so 
    # the model starts as an identity function baseline.
    delta = nn.Dense(36, kernel_init=nn.initializers.zeros)(h)

    # Autoregressive update: y_t = y_{t-1} + delta
    next_state = past_pred + delta

    # Ensure valid range [0, 1]
    return jnp.clip(next_state, 0.0, 1.0)


transition_model = Candidate()  # NOTE: Do keep this line as-is.


#
# OPTIMIZER
#

# Learning rate schedule with warmup and cosine decay for stability
lr_schedule = optax.warmup_cosine_decay_schedule(
    init_value=1e-6,
    peak_value=LEARNING_RATE,
    warmup_steps=2000,
    decay_steps=MAX_STEPS,
)

optimizer = optax.adamw(learning_rate=lr_schedule, weight_decay=1e-4)
\end{pythonbox}

\subsection{Solution \model{ts}[422]}
\label{app:ts_422}

\begin{pythonbox}
import jax
import jax.numpy as jnp
import flax.linen as nn
import optax

#
# CONSTANTS
#

BATCH_SIZE = 512
MAX_STEPS = 500_000
# Scheduled sampling enforces autoregressive stability.
# We decay slightly slower to allow the model to learn the base rules first.
SCHEDULED_SAMPLING_START_PROB = 1.0
SCHEDULED_SAMPLING_FINAL_PROB = 0.05
SCHEDULED_SAMPLING_DECAY_STEPS = 300_000
EARLY_STOP_PATIENCE = 1000
MIN_DELTA = 1e-9
LEARNING_RATE = 8e-4

#
# MODEL
#

class ResidualBlock(nn.Module):
    features: int

    @nn.compact
    def __call__(self, x, context):
        # Pre-norm architecture for stability
        norm = nn.LayerNorm()(x)
        
        # Gated Linear Unit (GEGLU variant)
        # Use context to modulate the hidden representation
        h = nn.Dense(self.features * 2, kernel_init=nn.initializers.orthogonal())(norm)
        h_val, h_gate = jnp.split(h, 2, axis=-1)
        h = h_val * nn.gelu(h_gate)
        
        # FiLM modulation using context
        gamma = nn.Dense(self.features, kernel_init=nn.initializers.zeros)(context)
        beta = nn.Dense(self.features, kernel_init=nn.initializers.zeros)(context)
        h = h * (1.0 + gamma) + beta
        
        h = nn.Dense(self.features, kernel_init=nn.initializers.orthogonal())(h)
        return x + h

class Candidate(nn.Module):
    """
    Refined Gated Dynamics Model.
    Models neural transitions using learned time-constants and target states,
    heavily conditioned on the behavioral 'Bout' switch.
    """
    
    @nn.compact
    def __call__(self, past_pred, cov_t, train=False):
        B, N = past_pred.shape # N=36
        
        # 1. Input Processing
        # Split covariates: Bout State (switch) and Visual Input (driving signal)
        bout = cov_t[:, 0:1]    
        visual = cov_t[:, 1:]   
        
        # High-dimensional visual embedding
        v_enc = nn.Dense(128)(visual)
        v_enc = nn.gelu(v_enc)
        
        # Context incorporates the Bout switch explicitly
        # We use the bout flag to gate visual information
        ctx = jnp.concatenate([
            v_enc, 
            bout, 
            v_enc * bout, # Bout-specific visual processing
            v_enc * (1.0 - bout) # Non-bout specific visual processing
        ], axis=-1)
        ctx = nn.Dense(128)(ctx)
        ctx = nn.gelu(ctx)
        
        # 2. Latent Projection
        # Project current neural state to a latent space
        latent_dim = 256
        x = nn.Dense(latent_dim, kernel_init=nn.initializers.orthogonal())(past_pred)
        
        # Apply N residual blocks with conditional modulation
        for i in range(3):
            x = ResidualBlock(features=latent_dim, name=f"res_block_{i}")(x, ctx)
        
        # 3. Liquid-Style Transition
        # We predict a 'gate' (alpha) and a 'target' state.
        # This ensures the output is always a convex combination of previous state
        # and a target, guaranteeing stability in [0, 1].
        
        # Gate: How much the state changes (1/tau)
        # We initialize it to be small to favor stability early on
        gate_logit = nn.Dense(N, kernel_init=nn.initializers.zeros)(x)
        alpha = nn.sigmoid(gate_logit - 3.0) 
        
        # Target: What the neurons are striving towards
        target = nn.Dense(N, kernel_init=nn.initializers.orthogonal())(x)
        target = nn.sigmoid(target) # Ensure target is in [0, 1]
        
        # Direct Delta: Small additive term for fast transient dynamics
        # This helps capture sudden shifts that a convex combination might miss.
        delta = nn.Dense(N, kernel_init=nn.initializers.zeros)(x)
        delta = jnp.tanh(delta) * 0.05
        
        # Update rule: y_t = (1-alpha)*y_{t-1} + alpha*target + delta
        next_state = (1.0 - alpha) * past_pred + alpha * target + delta
        
        return jnp.clip(next_state, 0.0, 1.0)

transition_model = Candidate() # NOTE: Do keep this line as-is.

#
# OPTIMIZER
#

lr_schedule = optax.warmup_cosine_decay_schedule(
    init_value=1e-7,
    peak_value=LEARNING_RATE,
    warmup_steps=4000,
    decay_steps=MAX_STEPS,
    end_value=1e-6
)

optimizer = optax.chain(
    optax.clip_by_global_norm(0.5), # Tighter clipping for autoregressive stability
    optax.adamw(
        learning_rate=lr_schedule,
        weight_decay=2e-3, # Increased weight decay for OOD robustness
        b1=0.9,
        b2=0.999
    )
)
\end{pythonbox}

\subsection{Solution \model{sts}[445]}
\label{app:sts_445}

\begin{pythonbox}
import jax
import jax.numpy as jnp
import flax.linen as nn
import optax

#
# CONSTANTS
#

BATCH_SIZE = 512
MAX_STEPS = 500_000
SCHEDULED_SAMPLING_START_PROB = 1.0
SCHEDULED_SAMPLING_FINAL_PROB = 0.0
SCHEDULED_SAMPLING_DECAY_STEPS = 400_000
EARLY_STOP_PATIENCE = 1500
MIN_DELTA = 1e-10
LEARNING_RATE = 1e-3

# NOTE: Connectivity from EPT (cols) to LPT (rows)
CONNECTIVITY = [
 [False, True, True, True, True, True, True, True],
 [False, True, True, True, True, False, True, False],
 [True, True, True, False, True, True, True, True],
 [True, True, True, True, True, False, True, False],
 [True, False, True, True, True, False, False, False],
 [False, False, True, False, True, False, False, False],
 [False, True, True, False, False, False, False, True],
 [False, True, True, True, False, True, True, True],
 [False, True, True, True, False, False, True, True],
 [False, False, False, False, True, True, False, False],
 [False, True, True, True, True, True, False, True],
 [False, True, True, True, True, True, False, True],
 [True, True, True, True, False, True, True, True],
 [True, False, True, False, False, True, True, True],
 [True, True, True, True, True, True, True, False],
 [True, False, True, False, True, True, True, True],
 [True, False, False, False, True, False, True, True],
 [True, False, False, False, False, False, True, False],
 [False, False, False, True, False, True, True, False],
 [False, True, True, True, False, True, True, True],
 [False, False, True, True, False, True, True, True],
 [True, True, False, False, False, False, False, False],
 [True, True, False, True, False, True, True, True],
 [True, True, False, True, False, True, True, True]]

#
# MODEL
#

class Candidate(nn.Module):
    """Hierarchical Neural Dynamic Model with Sparse Connectivity."""

    @nn.compact
    def __call__(self, past_pred, cov_t, train=False):
        B = past_pred.shape[0]
        
        # 1. Decomposition
        ept_past = past_pred[:, 0:8]     # (B, 8)
        ahb_past = past_pred[:, 32:36]   # (B, 4)
        bout_state = cov_t[:, 0:1]       # (B, 1)
        visual_in = cov_t[:, 1:9]        # (B, 8)

        # 2. EPT Update (Neurons 0-7)
        # Shared dynamics rule across all 8 neurons for OOD robustness.
        # Inputs: past state, specific visual channel, and global bout state.
        ept_combined = jnp.stack([ept_past, visual_in, jnp.broadcast_to(bout_state, (B, 8))], axis=-1)
        
        # Capture second-order interactions [y, v, b, y*v, y*b, v*b]
        ept_feat = jnp.concatenate([
            ept_combined,
            ept_combined[..., [0]] * ept_combined[..., [1]],
            ept_combined[..., [0]] * ept_combined[..., [2]],
            ept_combined[..., [1]] * ept_combined[..., [2]]
        ], axis=-1) # (B, 8, 6)

        # Shared MLP rule
        e = nn.Dense(128)(ept_feat)
        e = nn.leaky_relu(e)
        e = nn.Dense(128)(e)
        e = nn.leaky_relu(e)
        
        # Gated leaky integration for autoregressive stability
        ept_gate = nn.sigmoid(nn.Dense(1)(e)).squeeze(-1)
        ept_candidate = nn.sigmoid(nn.Dense(1)(e)).squeeze(-1)
        
        # Per-neuron individual gain/bias to capture phenotypic variation
        ept_gain = self.param('ept_gain', nn.initializers.ones, (8,))
        ept_out = (1.0 - ept_gate) * ept_past + ept_gate * (ept_candidate * ept_gain)

        # 3. LPT Update (Neurons 8-31)
        # Feed-forward mapping from current EPT, enforcing sparse CONNECTIVITY.
        mask = jnp.array(CONNECTIVITY) # (24 LPT rows, 8 EPT cols)
        
        # Two-layer sparse projection
        lpt_w1 = self.param('lpt_w1', nn.initializers.glorot_uniform(), (24, 8, 32))
        lpt_b1 = self.param('lpt_b1', nn.initializers.zeros, (24, 32))
        
        # Each LPT neuron integrates its subset of EPT inputs into a latent space
        lpt_h = jnp.einsum('bi,nik->bnk', ept_out, lpt_w1 * mask[..., None]) + lpt_b1
        lpt_h = nn.leaky_relu(lpt_h)
        
        lpt_w2 = self.param('lpt_w2', nn.initializers.glorot_uniform(), (24, 32, 1))
        lpt_b2 = self.param('lpt_b2', nn.initializers.zeros, (24,))
        lpt_out = jnp.einsum('bnk,nko->bn', lpt_h, lpt_w2) + lpt_b2
        lpt_out = nn.sigmoid(lpt_out)

        # 4. aHB/nMLF Update (Neurons 32-35)
        # Temporal integrators depending on past state and newly computed LPT.
        ahb_in = jnp.concatenate([ahb_past, lpt_out], axis=-1) # (B, 4 + 24)
        
        a = nn.Dense(256)(ahb_in)
        a = nn.leaky_relu(a)
        a = nn.Dense(256)(a)
        a = nn.leaky_relu(a)
        
        ahb_gate = nn.sigmoid(nn.Dense(4)(a))
        ahb_candidate = nn.sigmoid(nn.Dense(4)(a))
        ahb_out = (1.0 - ahb_gate) * ahb_past + ahb_gate * ahb_candidate

        return jnp.concatenate([ept_out, lpt_out, ahb_out], axis=-1)

transition_model = Candidate()

#
# OPTIMIZER
#

lr_schedule = optax.cosine_onecycle_schedule(
    transition_steps=MAX_STEPS,
    peak_value=LEARNING_RATE,
    pct_start=0.1,
    div_factor=10,
    final_div_factor=100
)

optimizer = optax.adamw(
    learning_rate=lr_schedule,
    weight_decay=1e-5
)
\end{pythonbox}

\newpage
\clearpage

\section{Supplementary Figures}
\label{app:supp_figs}

\begin{figure*}[h]
  \begin{center}
    \centerline{\includegraphics[width=\linewidth]{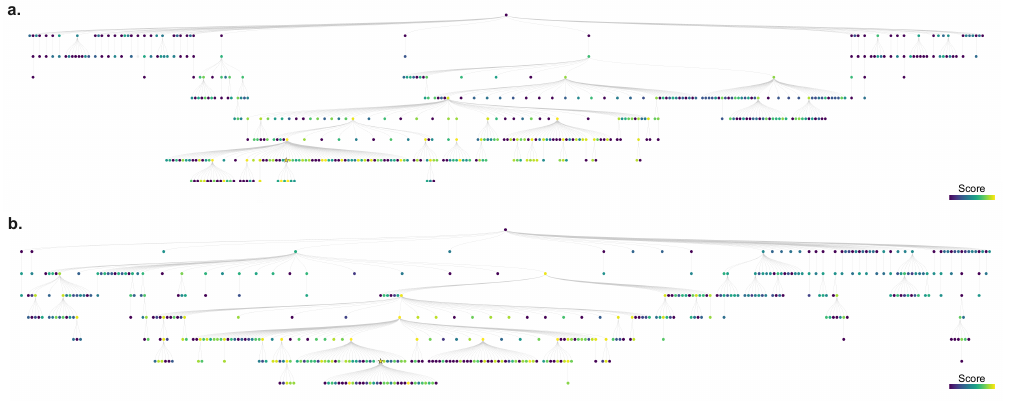}}
    \caption{
      \textbf{Full search trees for automated model discovery.}
      Visualization of the complete evolutionary lineage explored by the LLM-guided tree search. Nodes represent executable Python code (candidate models), and edges represent mutations generated by the LLM. Node color indicates the score, with lighter colors representing lower error.
      \textbf{a.} The unconstrained search tree (\model{ts}).
      \textbf{b.} The structure-constrained search tree (\model{sts}).
    }
    \label{fig:trees}
  \end{center}
\end{figure*}

\begin{figure*}[h]
  \begin{center}
    \centerline{\includegraphics[width=\linewidth]{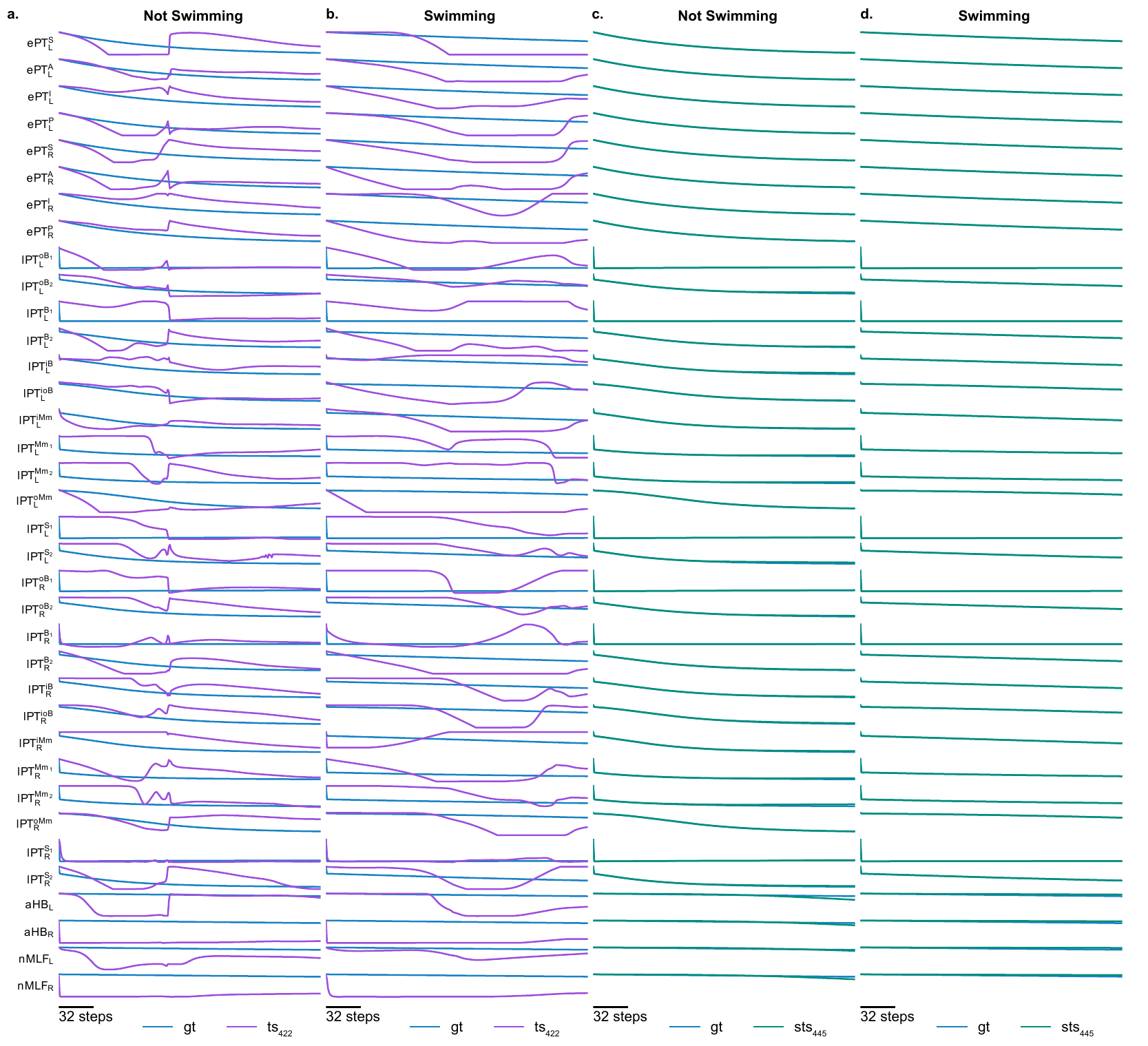}}
    \caption{
      \textbf{Impulse response dynamics.}
      We probe the recovery of dynamic mechanisms by the discovered models by initializing all neurons to a saturated state ($\mathbf{a}_0 = \mathbf{1}$) and evolving the system for 32 steps under ``Not Swimming'' (Rest) and ``Swimming'' conditions.
      \textbf{a, b.}~\model{ts}[422] (purple) exhibits erroneous oscillatory behavior, deviating significantly from the ground truth (blue).
      \textbf{c, d.}~\model{sts}[445] (green) faithfully reproduces the ground truth state-dependent decay profiles that distinguish swimming from resting dynamics.
    }
    \label{fig:impulse_responses}
  \end{center}
\end{figure*}

\newpage
\clearpage

\section{Supplementary Tables}
\label{app:metrics}

\begin{table}[h!]
\caption{\textbf{Performance comparison between best-scoring tree search solutions.} In addition to predictive accuracy (MAE), we report mechanistic verification metrics defined in \cref{app:additional_analyses}: the Frobenius norm of the effective connectivity mismatch ($\mathcal{L}_{\text{Jac}}$) and the impulse response error ($\mathcal{L}_{\text{IR}}$) for both resting and swimming states. The structure-constrained solution (\model{sts}[445]) demonstrates superior mechanistic fidelity compared to the unconstrained one (\model{ts}[422]).}
\label{tab:results_best}
\begin{center}
\begin{tabular}{lcc}
\toprule
 & \model{ts}[422] & \model{sts}[445] \\
\midrule
Test MAE $\downarrow$ & 0.00005 & \textbf{0.00001} \\
Holdout MAE $\downarrow$ & 0.13995 & \textbf{0.00027} \\
$\mathcal{L}_{\text{Jac}}$ $\downarrow$ & 8.33 & \textbf{0.26} \\
$\mathcal{L}_{\text{IR}}^{\text{rest}}$ $\downarrow$ & 0.27 & \textbf{0.04} \\
$\mathcal{L}_{\text{IR}}^{\text{swim}}$ $\downarrow$ & 0.29 & \textbf{0.04} \\
\bottomrule
\end{tabular}
\end{center}
\end{table}

\begin{table}[h!]
\caption{\textbf{Population statistics (median [IQR]) for the top 50 highest-scoring models of each tree search.} Aggregated metrics across models discovered during the unconstrained (\model{ts}) and structure-constrained (\model{sts}) tree searches. The constrained search consistently yields models with lower Jacobian error ($\mathcal{L}_{\text{jac}}$) and more accurate dynamics ($\mathcal{L}_{\text{IR}}$), indicating that the structural prior effectively regularizes the search space towards higher mechanistic fidelity.}  
\label{tab:results_median}
\begin{center}
\begin{tabular}{lcc}
\toprule
 & \model{ts} & \model{sts} \\
\midrule
Test MAE $\downarrow$ & 0.00009 [0.00008, 0.00009] & \textbf{0.00004 [0.00002, 0.00005]} \\
Holdout MAE $\downarrow$ & 0.00833 [0.00636, 0.02349] & \textbf{0.00030 [0.00025, 0.00073]} \\
$\mathcal{L}_{\text{Jac}}$ $\downarrow$ & 7.58 [7.55, 7.73] & \textbf{0.34 [0.23, 0.45]} \\
$\mathcal{L}_{\text{IR}}^{\text{rest}}$ $\downarrow$ & 0.14 [0.10, 0.21] & \textbf{0.03 [0.02, 0.04]} \\
$\mathcal{L}_{\text{IR}}^{\text{swim}}$ $\downarrow$ & 0.12 [0.09, 0.18] & \textbf{0.03 [0.02, 0.04]} \\
\bottomrule
\end{tabular}
\end{center}
\end{table}

\newpage

\clearpage
\refstepcounter{section}
\renewcommand{\thesection}{}
\addcontentsline{toc}{section}{\protect\numberline{\thesection}{References}}

\bibliographyapp{main}
\bibliographystyleapp{abbrvnat}

\end{document}